\documentclass[11pt,a4paper]{article}
\usepackage[utf8]{inputenc}
\usepackage[pagewise]{lineno}

\usepackage{epic,eepic,epsfig,amsmath,amssymb,amsthm,mathabx}
\usepackage{t1enc,tabularx,subcaption,graphicx}
\usepackage{array}
\usepackage{multirow}
\usepackage{blindtext}

\usepackage[francais,english]{babel}

\usepackage{caption}

\usepackage{color}

\usepackage{bbm}

\usepackage{hyperref}

\def\virgp{\raise 2pt\hbox{,}}

\renewcommand{\geq}{\geqslant}
\renewcommand{\leq}{\leqslant}

\def\R{{\mathbb R}}

\def\virgp{\raise 2pt\hbox{,}}
\def\cdotpv{\raise 2pt\hbox{;}}

\def\1{\mathbbm{1}}

\newtheorem{theorem}{Theorem}[section]
\newtheorem{corollary}[theorem]{Corollary}
\newtheorem{proposition}[theorem]{Proposition}
\newtheorem{lemma}[theorem]{Lemma}

\theoremstyle{remark}
\newtheorem{remark}{Remark}[section]

\theoremstyle{definition}
\newtheorem{definition}{Definition}[section]

\theoremstyle{definition}

\theoremstyle{definition}

\newtheorem{conjecture}[theorem]{Conjecture}

\newtheorem{assumption}[theorem]{Assumption}

\newcolumntype{M}[1]{>{\centering}m{#1}}

\begin{document}

\title{Walrasian \emph{non-t\^atonnement}: a kinetic model of bargaining on an Eid souk}
 
\author{Nizar Riane$^\dag$ and Claire David$^\ddag$}

\maketitle

\centerline{Mohammed V University\footnote{Nizar.Riane@gmail.com}}
\centerline{FSJESR -- Agdal, Laboratoire d'Économie Appliquée, Rabat, Morocco}

\vskip 0.25cm

\centerline{$^\ddag$ Sorbonne Universit\'e \footnote{Claire.David@sorbonne-universite.fr}}

\centerline{CNRS, UMR 7598, Laboratoire Jacques-Louis Lions, 4, place Jussieu 75005, Paris, France}
\vskip 0.5cm

\begin{abstract}
We propose a kinetic model of price formation in bargaining markets, inspired by the sheep souks held in Morocco before Eid al-Adha. Buyers and sellers, described by their current prices and their reservation prices, meet at random: each meeting either ends with a trade, after which both agents leave the market, or with a separation, after which both agents have conceded part of the gap between their prices. The resulting Boltzmann-type equation with losses is well-posed in the space of measures. We prove that the market converges to an exhausted state, in which no mutually profitable trade remains, and that a price bracket separates the remaining buyers and sellers, which provides a decentralized counterpart of Marshall's scissors. Numerical simulations show that this bracket lies close to the Walrasian price, while decentralized bargaining produces more trades than the Walrasian allocation, and transaction prices depend on the bargaining power of each side and on the timing of supply shocks.
\end{abstract}

\maketitle

\vskip 1cm


\vskip 1cm


\vskip 1cm


\vskip 0.5cm
 
\tableofcontents

\vskip 0.5cm

\section{Introduction}

\hskip 0.5cm In 1998, in response to a request from Vladimir Arnold, then vice-president of the International Mathematical Union, Steve Smale published a list of mathematical problems for the next century \cite{Smale1998}. His eighth problem asks for the introduction of dynamics into economic theory, and in particular for an extension of the mathematical model of general equilibrium that includes price adjustment.\\

The theory of general equilibrium was introduced by L\'eon Walras \cite{Walras1874}, and reformulated in the static case by Lionel McKenzie, Kenneth Arrow and Gerard Debreu \cite{Debreu1959}. Dynamics was introduced through the excess demand function $Z(p)$ and the t\^atonnement equation $\dot{p}=Z(p)$ \cite{Samuelson1983,Arrow1958,Arrow1959}: the price of a good rises when demand exceeds supply, and falls otherwise. In this description, prices adjust before any exchange takes place, and no mechanism explains who changes the prices, and how. Moreover, the t\^atonnement may fail to converge: Scarf gave examples in which prices oscillate around the equilibrium without approaching it \cite{Scarf1960}. In the theory of partial equilibrium, which studies a single market, the dominant picture remains Marshall's pair of scissors: the equilibrium is the price at which the demand and supply curves cross. Here again, the curves are postulated, and the process which leads the market to their intersection is not described. These theories raise questions about the existence, uniqueness and stability of equilibrium, which are usually answered by imposing restrictions on the demand and supply functions, without describing the mechanisms behind them.\\

Two families of models depart from the Walrasian t\^atonnement. In \emph{non-t\^atonnement} processes, trades take place out of equilibrium, while prices adjust \cite{HahnNegishi1962,Uzawa1962}. In the barter process of Edgeworth, formalized by Uzawa \cite{Uzawa1962}, agents trade as long as a mutually profitable exchange exists, and the final allocation depends on the path of trades, and not only on the initial endowments; in the process of Hahn and Negishi \cite{HahnNegishi1962}, trades are such that no agent is rationed on both sides of the market. In these models, however, prices still follow an aggregate law of the form $\dot{p}=Z(p)$, and Fisher noted that such a law describes the behaviour of no actual agent; he proposed models in which prices are set by the agents themselves \cite{Fisher1972}, and argued that, without a theory of disequilibrium behaviour, the practice of assuming equilibrium has no foundation \cite{Fisher1983}. In \emph{non-Walrasian} equilibrium theory, initiated by Clower \cite{Clower1965} and developed by Barro and Grossman \cite{BarroGrossman1971}, B\'enassy \cite{Benassy1975} and Dr\`eze \cite{Dreze1975}, prices are rigid in the short run, and markets are cleared by quantity constraints, that is, by rationing.\\

Real markets suggest a different picture, in which prices are formed through bilateral interactions and exchanges take place out of equilibrium. A typical example is the sheep market held every week in Morocco before Eid al-Adha. Buyers arrive at the souk with an opening offer and a maximal price, set by their budget and preferences; sellers arrive with an asking price, set by their breeding costs and the profit they hope for, and a minimal price below which they do not sell. A buyer asks a seller for his price, and the meeting becomes a bargaining session: each side concedes part of the gap between the two prices. If an agreement is reached, the sheep is sold and both agents leave the market; otherwise, the buyer moves on to another seller, now ready to pay a little more, and the seller is ready to accept a little less. Such bazaar economies, in which information is scarce and prices are bargained, were described by Geertz on the souk of Sefrou \cite{Geertz1978}. Nothing guarantees that a common price exists at the beginning of the day, nor that all trades take place at the same price. Although we use the Eid souk as a guiding example, the same description applies to bargaining markets in general, such as bazaars, livestock markets, flea markets or second-hand markets between individuals, where each agent buys or sells one unit and prices are negotiated bilaterally.\\

Decentralized markets of this kind have also been studied experimentally. Chamberlin organized decentralized one-shot markets in which students bargained in pairs, and found that the number of trades was typically larger, and the prices typically lower, than predicted by the competitive equilibrium \cite{Chamberlin1948}; he could account for the excess of trades, but the lower prices remained unexplained. Smith then showed that centralized and repeated trading, in double auctions, brings prices close to the equilibrium \cite{Smith1962}, which changes the institution rather than explaining the bias of the decentralized market. Gode and Sunder showed that markets populated by agents with random bids, constrained only by their reservation prices, reach almost full allocative efficiency \cite{GodeSunder1993}. In a closely related spirit, Inoua and Smith formalize a classical view of supply and demand in which the reservation prices of buyers and sellers are the primitive data and competition is a process of higgling and bargaining, and argue that the early experimental markets support this view rather than neoclassical price theory \cite{InouaSmith2022}; in their theory, the competitive price minimizes the surplus of mutually beneficial trades left unrealized, which decreases along the sequence of transaction prices. On the theoretical side, decentralized markets with sequential bargaining between rational agents were studied in \cite{Rubinstein1985,Gale1987}, and Satterthwaite and Shneyerov proved that, in a dynamic matching market in which buyers and sellers leave after trading, prices converge to the Walrasian price as frictions vanish \cite{SatterthwaiteShneyerov2007}. Empirically, the wholesale fish market of Marseille has been studied in detail, and an agent-based model of this market explains both price dispersion and the loyalty of buyers \cite{KirmanVriend2001}.\\

These bilateral meetings are also analogous to collisions between particles. In the Boltzmann equation, one does not follow the position and the velocity of each particle, but the distribution of a typical particle, which evolves under three effects: external forces, diffusion, and binary collisions \cite{Cercignani1994,Villani2002}. Only the last effect is relevant here: a meeting between a buyer and a seller is a collision, which changes their prices. Kinetic equations have been widely used to describe interacting agents in economics and social sciences \cite{Pareschi2013}: Boudin and Salvarani derived a kinetic model of opinion formation, in which two individuals exchange opinions with respect to their average value before the interaction \cite{Boudin2009}, and kinetic models of markets and price formation have been proposed in \cite{Cordier2005,LasryLions2007,Burger2013}. In the Boltzmann-type model of Burger et al. \cite{Burger2013}, a buyer and a vendor who trade exchange their roles, so that the numbers of buyers and vendors are conserved, as in financial markets. In a souk, on the contrary, an agent who has traded leaves the market.\\

In this paper, we propose a kinetic model of bargaining markets, with the Eid souk as a guiding example. Each buyer is described by his offer and his reservation price, each seller by his asking price and his cost. A meeting either ends with a trade, after which both agents leave the market, or with a separation, after which both agents have conceded part of the gap. The supply and demand curves are not postulated: they are given by the distributions of the reservation prices. Our model lies between the two families described above: as in non-t\^atonnement processes, trades take place out of equilibrium and the final state depends on the path of the market; as in non-Walrasian equilibria, some agents are rationed, since they leave the souk without trading; but, unlike both, prices are neither given by an aggregate law nor fixed, and they are bargained by the agents in each meeting. Unlike the models of \cite{Rubinstein1985,Gale1987,SatterthwaiteShneyerov2007}, our agents are not strategic: as in \cite{GodeSunder1993}, they follow simple bargaining rules, bounded by their reservation prices, and we study the transient dynamics of a single day rather than a stationary state. We share the classical starting point of \cite{InouaSmith2022}, and provide an explicit bargaining dynamics, in which agents leave the market after trading, together with proofs of convergence. This framework recovers, in simulations, the two observations of Chamberlin: more trades than at the competitive equilibrium, and lower prices, for which we propose an explanation by a selection effect: the sellers with low costs announce low prices and sell first. Our main results are the following.\\

\begin{enumerate}
\item[\emph{i}.] In Section \ref{Section1}, we describe the agents and the bargaining rule, and we derive a Boltzmann-type equation with losses, in which buyers and sellers leave the market by pairs.\\
\item[\emph{ii}.] In Section \ref{Section2}, we prove that this equation is well-posed in the space of measures: it has a unique nonnegative solution.\\
\item[\emph{iii}.] In Section \ref{Section3}, we prove that the market converges to an \emph{exhausted state}, in which no mutually profitable trade remains, and we show that a price bracket separates the remaining buyers and sellers, which provides a decentralized counterpart of Marshall's scissors. These results extend to agents whose expectations vary during the day, or respond to the number of agents still present, and to agents who buy or sell several sheep.\\
\item[\emph{iv}.] In Section \ref{Section4}, we simulate the souk with a particle method. The final bracket lies close to the Walrasian price, decentralized bargaining produces more trades than the Walrasian allocation, and the transaction prices depend on the bargaining power of each side, on the timing of supply shocks, and on the expectations of the agents.\\
\end{enumerate}

\vskip 0.5cm

\section{The Eid sheep market model\label{Section1}}

\hskip 0.5cm We consider a single good, sheep, sold on a weekly souk before Eid al-Adha, under the following assumptions:
\begin{enumerate}
\item a large number of buyers and sellers;
\item homogeneous products: sheep differ only through the prices of their buyers and sellers;
\item buyers and sellers interact through binary meetings, which are brief events, localized in time;
\item meetings are inelastic: they reduce the bargaining room of the agents, or end with a trade;
\item Boltzmann's chaos assumption: the states of a buyer and of a seller who are about to meet are uncorrelated.
\end{enumerate}
One souk day is represented by a time interval $[0,T]$. Each buyer wants exactly one sheep and each seller brings exactly one sheep; a buyer or a seller with several sheep can be represented by several buyers or sellers. This representation neglects the correlations between the sheep of a same agent, who usually offers or asks similar prices for similar sheep and adjusts them jointly; agents who trade several sheep, described by their stock, are considered in Section \ref{Section3}.

\subsection{Agents}

\hskip 0.5cm A \emph{buyer} is described by a pair
\begin{align*}
(x,v)\in\Omega_b:=\left\lbrace (x,v)\in(\R_+)^2 \, , \, x\leq v \right\rbrace ,
\end{align*}
where $v$ is his \emph{reservation price}, i.e.\ the maximal price he is ready to pay, which summarizes his budget and preferences, and $x$ is his current offer.\\

A \emph{seller} is described by a pair
\begin{align*}
(y,c)\in\Omega_s:=\left\lbrace (y,c)\in(\R_+)^2 \, , \, y\geq c \right\rbrace ,
\end{align*}
where $c$ is his reservation price, i.e.\ the minimal price he accepts, which summarizes his breeding costs, and $y$ is his current asking price.\\

Let $f(t)$ (resp. $g(t)$) be the distribution of buyers on $\Omega_b$ (resp. of sellers on $\Omega_s$) at time $t\geq 0$: it is a nonnegative measure, and, for a Borel set $A\subset\Omega_b$, $f(t)(A)$ is the number of buyers whose state $(x,v)$ lies in $A$. Their total masses
\begin{align*}
N_b(t)=f(t)(\Omega_b) \quad , \quad N_s(t)=g(t)(\Omega_s)
\end{align*}
are the numbers of buyers and sellers still present on the market at time $t$. Since the cap introduced below concentrates buyers on the diagonal $\{x=v\}$ and sellers on $\{y=c\}$, $f(t)$ and $g(t)$ are in general measures, and not functions; we write $\int\phi\,df(t)$ for the integral of a function $\phi$ with respect to the measure $f(t)$.

\vskip 0.5cm

\begin{remark}[Supply and demand]{\ }\\
The demand and supply functions do not need to be postulated: the \emph{Marshall curves} are given by the initial distributions of reservation prices,
\begin{align}
\label{eqn:1}
D(p)=\int_{\Omega_b} \1_{\{v\geq p\}}\, df_0 \quad , \quad S(p)=\int_{\Omega_s} \1_{\{c\leq p\}}\, dg_0 \, ,
\end{align}
that is, $D(p)$ is the number of buyers ready to pay at least $p$, and $S(p)$ the number of sellers ready to sell at $p$. The function $D$ is nonincreasing and $S$ is nondecreasing. A price $e$ at which the curves $D$ and $S$ cross, i.e.\ $D(e)=S(e)$ when such a price exists, is the competitive equilibrium price of the souk \cite{Walras1874,Marshall1890}; we use it as a \emph{Walrasian benchmark}, to which the outcome of the bargaining is compared, and $D(e)$ is the corresponding \emph{Walrasian volume}.
\end{remark}

\vskip 0.5cm

\subsection{The bargaining rule}

\hskip 0.5cm When a buyer $(x,v)$ meets a seller $(y,c)$, the buyer asks the price and the seller announces $y$.
\begin{enumerate}
\item If $y\leq x$, the trade takes place at the price $y$, and both agents leave the market.
\item If $y>x$, the two agents bargain. Each of them concedes a fraction of the gap $y-x$:
\begin{align}
\label{eqn:2}
\begin{cases}
x'=\min\left(v \, , \, x+a\,(y-x)\right)\\
y'=\max\left(c \, , \, y-b\,(y-x)\right)
\end{cases}
\end{align}
where the concession rates $(a,b)\in[0,1]^2$ are drawn, independently for each meeting, from a probability measure $\Lambda$ on $[0,1]^2$.
\begin{enumerate}
\item If $x'\geq y'$, the trade takes place at the price $y'$, and both agents leave the market.
\item If $x'<y'$, the agents separate, the buyer with the new offer $x'$, the seller with the new asking price $y'$.
\end{enumerate}
\end{enumerate}

\vskip 0.5cm

\begin{remark}[Local information]{\ }\\
The driving force of rule (\ref{eqn:2}) is the price gap $y-x$, which both agents observe during the meeting. Each agent reacts only to this gap, and not to the state of the whole market, such as the number of sellers he has visited or the prices he has heard of; as in the zero-intelligence markets of \cite{GodeSunder1993}, the agents are not strategic. Aggregate supply and demand act indirectly, through the reservation prices $v$ and $c$, which bound the concessions, and through the selection of the agents who leave the market. The concession rates $a$ and $b$ measure the bargaining behaviour of the two sides; random rates make the outcome of a negotiation depend on the gap, and not only on the parameters. Concession rates which change during the day, for instance in response to the state of the market, are considered in Section \ref{Section3}.
\end{remark}

\vskip 0.5cm

\begin{lemma}\label{souk:lemma}{\ }\\
For every meeting:
\begin{enumerate}
\item if $y>x$, then $x\leq x'\leq v$ and $c\leq y'\leq y$: buyers only raise their offers, sellers only lower their asking prices, and both stay within their reservation prices. In particular, $(x',v)\in\Omega_b$ and $(y',c)\in\Omega_s$.
\item Every transaction price $p$ satisfies $c\leq p\leq v$. In particular, only pairs with $c\leq v$ can trade.
\end{enumerate}
\end{lemma}

\vskip 0.5cm

\begin{proof}
Since $x\leq v$ and $y>x$, $x+a(y-x)\geq x$, hence $x\leq x'\leq v$. The same argument gives $c\leq y'\leq y$. If $y\leq x$, the price is $p=y$ and $c\leq y\leq x\leq v$. If $y>x$ and $x'\geq y'$, the price is $p=y'$ and $c\leq y'\leq x'\leq v$.
\end{proof}

\vskip 0.5cm

Since $x'\leq x+a(y-x)$ and $y'\geq y-b(y-x)$, a trade after bargaining requires $a+b\geq 1$. When the support of $\Lambda$ lies in $\{a+b<1\}$, every trade takes place at the asking price of the seller, in a meeting with $y\leq x$, and the bargaining only brings the prices closer for future meetings.

\vskip 0.5cm

\subsection{The kinetic equation}

\hskip 0.5cm Meetings occur at random, at rate $\beta>0$ per buyer-seller pair, so that a buyer meets sellers at rate $\beta N_s(t)$: when the souk empties, meetings become rarer. We keep this simplest choice, which makes the collision operator quadratic. By the chaos assumption, the pairs which meet are distributed according to the product of $f$ and $g$. We denote by
\begin{align}
\label{eqn:3}
\chi(x,v,y,c,a,b)=\1_{\{y>x\}}\,\1_{\{y'>x'\}}
\end{align}
the indicator of a \emph{separation}, where $x'$ and $y'$ are given by (\ref{eqn:2}). A meeting ends with a trade if and only if $\chi=0$. To shorten the notation, we write
\begin{align*}
\pi_t = f(t)\otimes g(t)\otimes\Lambda \qquad \text{on } \ \Theta:=\Omega_b\times\Omega_s\times[0,1]^2 \, .
\end{align*}

For every bounded continuous test functions $\phi$ on $\Omega_b$ and $\psi$ on $\Omega_s$, the evolution of $f$ and $g$ is given, in weak form, by
\begin{align}
\label{eqn:4}
\begin{cases}
\displaystyle\frac{d}{dt}\int_{\Omega_b} \phi\, df(t) = \beta\int_{\Theta} \Big(\chi\,\phi(x',v)-\phi(x,v)\Big)\, d\pi_t \\ \\
\displaystyle\frac{d}{dt}\int_{\Omega_s} \psi\, dg(t) = \beta\int_{\Theta} \Big(\chi\,\psi(y',c)-\psi(y,c)\Big)\, d\pi_t
\end{cases}
\end{align}
with initial conditions $f(0)=f_0$ and $g(0)=g_0$. In words: every meeting removes the buyer from $(x,v)$, and only a separation puts him back, at $(x',v)$.\\

Splitting the integrand of the buyers' equation according to the three possible outcomes of a meeting, we obtain
\begin{align}
\label{eqn:5}
\chi\,\phi(x',v)-\phi(x,v) = \underbrace{-\,\1_{\{y\leq x\}}\,\phi(x,v)}_{\text{immediate trade}}\;
\underbrace{-\,\1_{\{y>x\}}\1_{\{x'\geq y'\}}\,\phi(x,v)}_{\text{trade after bargaining}}\;
+\;\underbrace{\1_{\{y>x\}}\1_{\{y'>x'\}}\big(\phi(x',v)-\phi(x,v)\big)}_{\text{separation}} \, .
\end{align}
The last term is a classical Boltzmann collision term (see \cite{Cercignani1994,Villani2002}), as in the opinion formation model of \cite{Boudin2009}: the buyer is moved from $x$ to $x'$. The first two terms are \emph{losses without gain}: a buyer who has bought a sheep leaves the market and does not reappear at any price.

\vskip 0.5cm

\begin{remark}[Why reservation prices are needed]{\ }\\
Consider the simpler rule $x'=x+\mu\,(y-x)$, $y'=y+\nu\,(x-y)$, with constant $\mu,\nu\in(0,1)$, without reservation prices and without trades, so that $f(t)$ and $g(t)$ are measures on $\R_+$, the sets of offers $x$ and of asking prices $y$, and $\chi=1$. Since $x'$ and $y'$ are convex combinations of $x$ and $y$, the prices remain in $[0,R]$ if they start there, and the masses $N_b$ and $N_s$ are conserved. Each meeting preserves $\nu x+\mu y$, since $\nu x'+\mu y'=\nu x+\mu y+\nu\mu\big((y-x)+(x-y)\big)$; taking $\phi(x)=x$ and $\psi(y)=y$ in the weak form, we obtain
\begin{align*}
\frac{d}{dt}\left(\nu\int x\,df(t)+\mu\int y\,dg(t)\right)=\beta\iint \big(\nu\mu\,(y-x)+\mu\nu\,(x-y)\big)\,df(t)(x)\,dg(t)(y)=0 \, .
\end{align*}
Consequently, if all prices converge to a common value $e_0$, then
\begin{align*}
e_0=\frac{\nu\int x\,df_0+\mu\int y\,dg_0}{\nu\,N_b(0)+\mu\,N_s(0)} \, ,
\end{align*}
a weighted average of the opening prices, which does not depend on supply and demand. The reservation prices and the exits introduced here are what bring supply and demand into the model.
\end{remark}

\vskip 0.5cm

\begin{remark}[Strong form]{\ }\\
Assume that $f(t)$ and $g(t)$ have densities, still denoted by $f(t,x,v)$ and $g(t,y,c)$. On the open set $\{x<v\}$, the cap in (\ref{eqn:2}) is inactive, and equation (\ref{eqn:4}) can be written $\partial_t f=\mathcal{Q}_b^+(f,g)-\beta\,N_s(t)\,f$, with the loss frequency $\beta N_s(t)$. A buyer arrives at the offer $x$ after a separation if his offer before the meeting was $x_*$, with $x=(1-a)\,x_*+a\,y$; inverting this relation at fixed $(y,a)$, with $a<1$, gives
\begin{align}
\label{eqn:6}
\mathcal{Q}_b^+(f,g)(t,x,v)=\beta\int_{\Omega_s}\int_{[0,1)\times[0,1]} \frac{1}{1-a}\, f(t,x_*,v)\,g(t,y,c)\,\1_{\{x_*\geq 0\}}\,\1_{\{y>x\}}\,\1_{\{x<y'_*\}} \, d\Lambda(a,b)\,dy\,dc \, ,
\end{align}
where $x_*=\dfrac{x-a\,y}{1-a}$ is the offer before the meeting and $y'_*=\max\left(c\, , \, y-b\,(y-x_*)\right)$ the asking price after it. The factor $\frac{1}{1-a}$ is the inverse of the Jacobian $\partial x/\partial x_*=1-a$, the partner's price $y$ being fixed in the integration. The condition $y>x_*$ for bargaining is equivalent to $y>x$, since $y-x=(1-a)(y-x_*)$, and $x<y'_*$ expresses that the meeting ends with a separation; the condition $x_*\leq v$ holds automatically, since $x_*\leq x<v$. Because of the cap $x'=\min(v,\cdot)$, the map $x_*\mapsto x'$ is not invertible on the whole of $\Omega_b$: buyers who reach their reservation price accumulate on the diagonal $\{x=v\}$, where the gain term is a singular measure. This is why we work with the weak form (\ref{eqn:4}) and with measure solutions, which require no inversion.
\end{remark}

\vskip 0.5cm

\begin{definition}[Transaction prices]{\ }\\
The \emph{measure of transaction prices} $h(t)$ is a nonnegative measure on $\R_+$ which records the prices of the trades realized on $[0,t]$: for $p_1\leq p_2$, $h(t)([p_1,p_2])$ is the number of trades realized before $t$ at a price in $[p_1,p_2]$. It is defined by $h(0)=0$ and, for every bounded continuous function $\theta$ on $\R_+$,
\begin{align}
\label{eqn:7}
\frac{d}{dt}\int_{\R_+} \theta\, dh(t) = \beta\int_{\Theta}\Big(\1_{\{y\leq x\}}\,\theta(y)+\1_{\{y>x\}}\1_{\{x'\geq y'\}}\,\theta(y')\Big)\, d\pi_t \, ,
\end{align}
where $\int\theta\,dh(t)$ denotes the integral of $\theta$ with respect to the measure $h(t)$. In words, each immediate trade adds a mass at the price $y$, each trade after bargaining a mass at the price $y'$, and separations add nothing. In particular, $h(t)$ is nondecreasing in $t$, its total mass $h(t)(\R_+)$ is the number of trades realized before $t$, and the distribution of the transaction prices is the probability measure $h(t)/h(t)(\R_+)$.
\end{definition}

\vskip 0.5cm

Together with the numbers $N_s(T)$ of unsold sheep and $N_b(T)$ of buyers leaving without a sheep, $h(T)$ provides the observable outputs of the model.

\vskip 0.5cm

\section{Existence and uniqueness\label{Section2}}

\hskip 0.5cm We denote by $\mathcal{M}(E)$ the space of finite signed Borel measures on a set $E$, endowed with the total variation norm $\|\cdot\|$, and by $\mathcal{M}_+(E)$ its cone of nonnegative measures. We work in the Banach space
\begin{align*}
X=\mathcal{M}(\Omega_b)\times\mathcal{M}(\Omega_s) \quad , \quad \|(f,g)\|_X=\|f\|+\|g\| \, .
\end{align*}

\subsection{The collision operator}

\hskip 0.5cm For $(f,g)\in X$, we define the gain terms $\mathcal{Q}_b^+(f,g)\in\mathcal{M}(\Omega_b)$ and $\mathcal{Q}_s^+(f,g)\in\mathcal{M}(\Omega_s)$ by
\begin{align}
\label{eqn:8}
\left\langle\mathcal{Q}_b^+(f,g),\phi\right\rangle=\beta\int_\Theta \chi\,\phi(x',v)\,d(f\otimes g\otimes\Lambda) \quad , \quad \left\langle\mathcal{Q}_s^+(f,g),\psi\right\rangle=\beta\int_\Theta \chi\,\psi(y',c)\,d(f\otimes g\otimes\Lambda) \, .
\end{align}
By Lemma \ref{souk:lemma}, the maps $(x,v,y,c,a,b)\mapsto(x',v)$ and $(x,v,y,c,a,b)\mapsto(y',c)$ take their values in $\Omega_b$ and $\Omega_s$; they are continuous and $\chi$ is a Borel function, so that $\mathcal{Q}_b^+(f,g)$ and $\mathcal{Q}_s^+(f,g)$ are the images of the measure $\beta\,\chi\,(f\otimes g\otimes\Lambda)$ by these maps. In particular, they are well-defined finite measures, which are nonnegative when $f$ and $g$ are. The collision operator is
\begin{align}
\label{eqn:9}
\mathcal{Q}(f,g)=\Big(\mathcal{Q}_b^+(f,g)-\beta\, g(\Omega_s)\, f \, , \, \mathcal{Q}_s^+(f,g)-\beta\, f(\Omega_b)\, g\Big) \, ,
\end{align}
and system (\ref{eqn:4}) reads $\partial_t(f,g)=\mathcal{Q}(f,g)$.

\vskip 0.5cm

\begin{lemma}\label{souk:lipschitz}{\ }\\
$\mathcal{Q}$ is quadratic, i.e.\ $\mathcal{Q}(F)=B(F,F)$ for a bilinear map $B$ on $X$, and, for all $F,G\in X$,
\begin{align*}
\|\mathcal{Q}(F)\|_X\leq 4\beta\,\|F\|_X^2 \quad , \quad \|\mathcal{Q}(F)-\mathcal{Q}(G)\|_X\leq 4\beta\left(\|F\|_X+\|G\|_X\right)\|F-G\|_X \, .
\end{align*}
\end{lemma}

\vskip 0.5cm

\begin{proof}
For $F_1=(f_1,g_1)$ and $F_2=(f_2,g_2)$ in $X$, let
\begin{align*}
B(F_1,F_2)=\Big(\mathcal{Q}_b^+(f_1,g_2)-\beta\, g_2(\Omega_s)\, f_1 \, , \, \mathcal{Q}_s^+(f_1,g_2)-\beta\, f_1(\Omega_b)\, g_2\Big) \, .
\end{align*}
The map $B$ is bilinear, since $\mathcal{Q}_b^+$ and $\mathcal{Q}_s^+$ are linear in each of their arguments, and $\mathcal{Q}(F)=B(F,F)$. We use three facts: the function $\chi$ is an indicator function, so that $|\chi|\leq 1$; for the total variation norm, $\|f_1\otimes g_2\otimes\Lambda\|=\|f_1\|\,\|g_2\|\,\|\Lambda\|=\|f_1\|\,\|g_2\|$, since $\Lambda$ is a probability measure; and the image of a measure by a measurable map has a smaller or equal total variation. Since $\mathcal{Q}_b^+(f_1,g_2)$ is the image of $\beta\,\chi\,(f_1\otimes g_2\otimes\Lambda)$ by the map $(x,v,y,c,a,b)\mapsto(x',v)$, we obtain
\begin{align*}
\|\mathcal{Q}_b^+(f_1,g_2)\|\leq\beta\,\|\chi\,(f_1\otimes g_2\otimes\Lambda)\|\leq\beta\,\|f_1\otimes g_2\otimes\Lambda\|=\beta\,\|f_1\|\,\|g_2\| \, .
\end{align*}
Moreover, since $|g_2(\Omega_s)|\leq\|g_2\|$,
\begin{align*}
\|\beta\,g_2(\Omega_s)\,f_1\|=\beta\,|g_2(\Omega_s)|\,\|f_1\|\leq\beta\,\|f_1\|\,\|g_2\| \, ,
\end{align*}
and the same bounds hold for the two terms of the sellers. Summing the four terms, and using $\|f_1\|\leq\|F_1\|_X$ and $\|g_2\|\leq\|F_2\|_X$,
\begin{align*}
\|B(F_1,F_2)\|_X\leq 4\beta\,\|f_1\|\,\|g_2\|\leq 4\beta\,\|F_1\|_X\,\|F_2\|_X \, .
\end{align*}
With $F_1=F_2=F$, this gives the first inequality. By bilinearity,
\begin{align*}
\mathcal{Q}(F)-\mathcal{Q}(G)=B(F,F)-B(G,F)+B(G,F)-B(G,G)=B(F-G,F)+B(G,F-G) \, ,
\end{align*}
hence
\begin{align*}
\|\mathcal{Q}(F)-\mathcal{Q}(G)\|_X\leq 4\beta\,\|F-G\|_X\,\|F\|_X+4\beta\,\|G\|_X\,\|F-G\|_X=4\beta\left(\|F\|_X+\|G\|_X\right)\|F-G\|_X \, .
\end{align*}
\end{proof}

\vskip 0.5cm

\subsection{Main result}

\vskip 0.5cm

\begin{proposition}[Pairwise exit]\label{souk:mass}{\ }\\
Let $(f,g)$ be a nonnegative solution of (\ref{eqn:4}), and let
\begin{align*}
\Gamma(t)=\beta\int_\Theta (1-\chi)\, d\pi_t\;\geq 0
\end{align*}
be the number of trades per unit of time. Then
\begin{align}
\label{eqn:10}
\frac{d N_b}{dt}=\frac{d N_s}{dt}=-\,\Gamma(t) \, .
\end{align}
Consequently, $N_b$ and $N_s$ are nonincreasing, and $N_b(t)-N_s(t)=N_b(0)-N_s(0)$ for all $t\geq 0$. Moreover, the total mass of $h(t)$ is $N_b(0)-N_b(t)$.
\end{proposition}

\vskip 0.5cm

\begin{proof}
The constant functions $\phi=1$ on $\Omega_b$ and $\psi=1$ on $\Omega_s$ are bounded and continuous, and $\int\phi\,df(t)=f(t)(\Omega_b)=N_b(t)$, $\int\psi\,dg(t)=g(t)(\Omega_s)=N_s(t)$. With these test functions, the integrand of both equations of (\ref{eqn:4}) is $\chi-1=-(1-\chi)$, hence
\begin{align*}
\frac{dN_b}{dt}=\frac{dN_s}{dt}=-\beta\int_\Theta(1-\chi)\,d\pi_t=-\Gamma(t) \, .
\end{align*}
Since $1-\chi\geq 0$ and $\pi_t$ is a nonnegative measure, $\Gamma(t)\geq 0$, so that $N_b$ and $N_s$ are nonincreasing, and $\frac{d}{dt}\big(N_b-N_s\big)=0$, so that $N_b(t)-N_s(t)=N_b(0)-N_s(0)$.\\

Every meeting ends in exactly one of the three outcomes of rule (\ref{eqn:2}), so that
\begin{align*}
1=\1_{\{y\leq x\}}+\1_{\{y>x\}}\1_{\{x'\geq y'\}}+\1_{\{y>x\}}\1_{\{y'>x'\}}=\1_{\{y\leq x\}}+\1_{\{y>x\}}\1_{\{x'\geq y'\}}+\chi \, .
\end{align*}
Hence, with $\theta=1$, the integrand of (\ref{eqn:7}) is $\1_{\{y\leq x\}}+\1_{\{y>x\}}\1_{\{x'\geq y'\}}=1-\chi$, and
\begin{align*}
\frac{d}{dt}\int_{\R_+}1\,dh(t)=\Gamma(t)=-\frac{dN_b}{dt} \, .
\end{align*}
Since $h(0)=0$, integrating in time shows that the total mass of the measure $h(t)$ is $\int_{\R_+}1\,dh(t)=N_b(0)-N_b(t)$.
\end{proof}

\vskip 0.5cm

Proposition \ref{souk:mass} replaces the conservation of mass of the classical Boltzmann equation: buyers and sellers leave the market by pairs.

\vskip 0.5cm

\begin{theorem}[Existence and uniqueness]\label{souk:wellposed}{\ }\\
Let $f_0\in\mathcal{M}_+(\Omega_b)$, $g_0\in\mathcal{M}_+(\Omega_s)$, and let $\Lambda$ be a probability measure on $[0,1]^2$. There exists a unique $(f,g)\in C^0\big(\R_+;X\big)$ which solves (\ref{eqn:4}) with initial data $(f_0,g_0)$. Moreover, $(f,g)\in C^1(\R_+;X)$, $f(t)$ and $g(t)$ are nonnegative for all $t\geq 0$, and
\begin{align*}
\|f(t)\|=N_b(t)\leq N_b(0) \quad , \quad \|g(t)\|=N_s(t)\leq N_s(0) \, .
\end{align*}
\end{theorem}

\vskip 0.5cm

\begin{proof}
\emph{Step 1: weak and strong solutions.} By (\ref{eqn:8}) and (\ref{eqn:9}), the right-hand side of (\ref{eqn:4}) is $\langle\mathcal{Q}(f,g),(\phi,\psi)\rangle$. Let $F=(f,g)\in C^0(\R_+;X)$ be a solution of (\ref{eqn:4}). By Lemma \ref{souk:lipschitz}, $s\mapsto\mathcal{Q}(F(s))$ is continuous with values in the Banach space $X$, so that its integral $\int_0^t\mathcal{Q}(F(s))\,ds$, defined as the limit in $X$ of Riemann sums, is well defined (see \cite{Dieudonne1960}). Since $\mu\mapsto\int\phi\,d\mu$ is a continuous linear form on $\mathcal{M}(\Omega_b)$ for every bounded continuous $\phi$, it commutes with this integral; integrating (\ref{eqn:4}) in time, $F(t)$ and $F_0+\int_0^t\mathcal{Q}(F(s))\,ds$ therefore have the same integrals against every bounded continuous test function. A finite measure on a metric space is determined by these integrals, hence $F(t)=F_0+\int_0^t\mathcal{Q}(F(s))\,ds$ in $X$, so that $F\in C^1(\R_+;X)$ and $F'=\mathcal{Q}(F)$. Conversely, such a strong solution solves (\ref{eqn:4}).\\

\emph{Step 2: local existence and uniqueness.} By Lemma \ref{souk:lipschitz}, $\mathcal{Q}$ is Lipschitz continuous on bounded sets of $X$. The Cauchy-Lipschitz theorem in the Banach space $X$ (see \cite{Dieudonne1960}) gives a unique maximal solution $F\in C^1([0,T^*);X)$ of $F'=\mathcal{Q}(F)$, $F(0)=F_0$, and $T^*<+\infty$ implies $\|F(t)\|_X\to+\infty$ as $t\to T^*$. By Step 1, this is also the unique solution of (\ref{eqn:4}) in $C^0([0,T^*);X)$.\\

\emph{Step 3: nonnegativity.} Let $n_b(t)=f(t)(\Omega_b)$ and $n_s(t)=g(t)(\Omega_s)$, which are continuous real functions, a priori of any sign. Since $\partial_t f+\beta\,n_s(t)\,f=\mathcal{Q}_b^+(f,g)$ is a linear equation in $X$ with a scalar coefficient, the solution satisfies the Duhamel formula
\begin{align}
\label{eqn:11}
f(t)=e^{-\beta\int_0^t n_s(\sigma)\,d\sigma}\,f_0+\int_0^t e^{-\beta\int_s^t n_s(\sigma)\,d\sigma}\,\mathcal{Q}_b^+\big(f(s),g(s)\big)\,ds \, ,
\end{align}
and similarly for $g$, with $n_b$ in place of $n_s$. Fix $t_0<T^*$, let $R=2\sup_{[0,t_0]}\|F\|_X$, and let $\tau=\min\left(t_0,1/(4\beta R)\right)$. With $n_b$ and $n_s$ \emph{fixed} (those of the solution), the right-hand side of (\ref{eqn:11}) and its analogue for $g$ define a map $\Psi$ on the closed ball $\mathcal{B}$ of radius $R$ of $C^0([0,\tau];X)$, endowed with the norm $\sup_{[0,\tau]}\|\cdot\|_X$. Since $|n_b|,|n_s|\leq R/2$, the exponential factors are positive and bounded by $e^{\beta\tau R/2}\leq e^{1/8}$. By the bounds on $\mathcal{Q}_b^+$ and $\mathcal{Q}_s^+$ established in the proof of Lemma \ref{souk:lipschitz}, $\|\mathcal{Q}_b^+(F)\|+\|\mathcal{Q}_s^+(F)\|\leq 2\beta\|f\|\,\|g\|\leq\beta R^2/2$ and $\|\mathcal{Q}_b^+(F_1)-\mathcal{Q}_b^+(F_2)\|+\|\mathcal{Q}_s^+(F_1)-\mathcal{Q}_s^+(F_2)\|\leq 2\beta R\,\|F_1-F_2\|_X$ for $F,F_1,F_2$ of norm at most $R$. Hence, for $F_1,F_2\in\mathcal{B}$,
\begin{align*}
\sup_{[0,\tau]}\|\Psi(F_1)\|_X\leq e^{1/8}\left(\frac{R}{2}+\tau\,\frac{\beta R^2}{2}\right)\leq\frac58\,e^{1/8}R<R \, , \qquad \sup_{[0,\tau]}\|\Psi(F_1)-\Psi(F_2)\|_X\leq 2\beta\tau R\,e^{1/8}\sup_{[0,\tau]}\|F_1-F_2\|_X \, ,
\end{align*}
with $2\beta\tau R\,e^{1/8}\leq\frac12e^{1/8}<1$: $\Psi$ maps $\mathcal{B}$ into itself and is a contraction on it. Its unique fixed point in $\mathcal{B}$ is the solution $F$, which belongs to $\mathcal{B}$ since $\sup_{[0,\tau]}\|F\|_X\leq R/2$, and it is the limit of the Picard iterates $\Psi^n(F_0)$. Since the exponential factors are positive and $\mathcal{Q}_b^+$ and $\mathcal{Q}_s^+$ map nonnegative measures to nonnegative measures, every iterate is nonnegative; as the cone of nonnegative measures is closed in $X$, $F$ is nonnegative on $[0,\tau]$. As $\tau$ depends only on $\beta$ and $R$, and $\|F(\tau)\|_X\leq R/2$, the argument can be repeated on $[\tau,2\tau]$, $[2\tau,3\tau]$, \dots, and $F$ is nonnegative on $[0,t_0]$, for every $t_0<T^*$.\\

\emph{Step 4: global existence.} Since $F$ is nonnegative on $[0,T^*)$, $\|f(t)\|=N_b(t)$ and $\|g(t)\|=N_s(t)$, and the proof of Proposition \ref{souk:mass}, which only differentiates these masses, applies on $[0,T^*)$: $N_b$ and $N_s$ are nonincreasing. Hence $\|F(t)\|_X\leq N_b(0)+N_s(0)$ on $[0,T^*)$, and $T^*=+\infty$ by Step 2.
\end{proof}

\vskip 0.5cm

\begin{proposition}[Bounded prices]\label{souk:support}{\ }\\
Let $R>0$, $\Omega_b^R=\Omega_b\cap\{v\leq R\}$ and $\Omega_s^R=\Omega_s\cap\{y\leq R\}$. We say that a measure $\mu$ on $\Omega_b$ (resp. $\Omega_s$) is concentrated on $\Omega_b^R$ (resp. $\Omega_s^R$) if $\mu\big(\Omega_b\setminus\Omega_b^R\big)=0$ (resp. $\mu\big(\Omega_s\setminus\Omega_s^R\big)=0$), i.e.\ if no buyer has a reservation price larger than $R$ (resp. no seller has an asking price larger than $R$). If $f_0$ and $g_0$ are concentrated on $\Omega_b^R$ and $\Omega_s^R$, then so are $f(t)$ and $g(t)$ for all $t\geq 0$.
\end{proposition}

\vskip 0.5cm

\begin{proof}
By Lemma \ref{souk:lemma}, $(x',v)\in\Omega_b^R$ when $(x,v)\in\Omega_b^R$, and $(y',c)\in\Omega_s^R$ when $(y,c)\in\Omega_s^R$. Hence $\mathcal{Q}$ maps the closed subspace $X^R=\mathcal{M}(\Omega_b^R)\times\mathcal{M}(\Omega_s^R)$ into itself, and Theorem \ref{souk:wellposed} applied in $X^R$ gives a solution with values in $X^R$, which is the solution by uniqueness.
\end{proof}

\vskip 0.5cm

\begin{remark}{\ }\\
The measure $\Lambda$ may depend on time, in order to model the end of the souk day, or the approach of Eid, when agents concede faster. If $t\mapsto\Lambda_t$ is piecewise continuous for the total variation norm, the proof above applies on each interval of continuity. In the next section, $\Lambda$ is fixed; laws depending on time, or on the numbers of agents still present, are considered in Section \ref{Section3}.
\end{remark}

\vskip 0.5cm

\section{Long-time behaviour\label{Section3}}

\hskip 0.5cm In this section, we study the state of the souk when the day is long compared with the time between two meetings, i.e.\ the limit $t\to+\infty$. We make the following assumptions.

\vskip 0.5cm

\begin{assumption}\label{souk:assumption}{\ }
\begin{enumerate}
\item[(A1)] There exists $R>0$ such that $f_0$ and $g_0$ are concentrated on $\Omega_b^R$ and $\Omega_s^R$.
\item[(A2)] The support of $\Lambda$ contains a point $(a,b)$ with $a>0$ and a point $(a,b)$ with $b>0$.
\end{enumerate}
\end{assumption}

\vskip 0.5cm

Assumption (A2) only excludes the degenerate cases where one side of the market never concedes.

\subsection{Lyapunov functionals}

\hskip 0.5cm We introduce the remaining \emph{bargaining room} of the buyers and of the sellers:
\begin{align}
\label{eqn:12}
E_b(t)=\int_{\Omega_b}(v-x)\,df(t) \quad , \quad E_s(t)=\int_{\Omega_s}(y-c)\,dg(t) \, .
\end{align}

\vskip 0.5cm

\begin{lemma}[Dissipation]\label{souk:dissipation}{\ }\\
Under Assumption (A1), $E_b$ and $E_s$ are nonincreasing, and
\begin{align}
\label{eqn:13}
\frac{dE_b}{dt}=-\,\mathcal{D}_b(t) \quad , \quad \frac{dE_s}{dt}=-\,\mathcal{D}_s(t) \, ,
\end{align}
with the nonnegative dissipations
\begin{align*}
\mathcal{D}_b(t)=\beta\int_\Theta\Big(\chi\,(x'-x)+(1-\chi)(v-x)\Big)\,d\pi_t \quad , \quad \mathcal{D}_s(t)=\beta\int_\Theta\Big(\chi\,(y-y')+(1-\chi)(y-c)\Big)\,d\pi_t \, .
\end{align*}
\end{lemma}

\vskip 0.5cm

\begin{proof}
By Proposition \ref{souk:support}, $f(t)$ is concentrated on $\Omega_b^R$, where $0\leq v-x\leq R$. Hence we can use $\phi(x,v)=\min(v-x,R)$, which is bounded and continuous on $\Omega_b$ and coincides with $v-x$ on $\Omega_b^R$, as a test function in (\ref{eqn:4}). The left-hand side is then $dE_b/dt$. Writing $v-x=\chi\,(v-x)+(1-\chi)(v-x)$, the integrand of the right-hand side reads
\begin{align*}
\chi\,(v-x')-(v-x)=-\chi\,(x'-x)-(1-\chi)\,(v-x) \, .
\end{align*}
The first term corresponds to a separation, in which the buyer raises his offer from $x$ to $x'$; the second one to a trade, in which the buyer leaves the market with his remaining bargaining room $v-x$. Both are nonpositive: $\chi\in\{0,1\}$, $x'\geq x$ when $\chi=1$ by Lemma \ref{souk:lemma}, and $v-x\geq 0$ on $\Omega_b$. This gives $dE_b/dt=-\mathcal{D}_b(t)$ with $\mathcal{D}_b(t)\geq 0$.\\

The same computation with $\psi(y,c)=\min(y-c,R)$ gives
\begin{align*}
\chi\,(y'-c)-(y-c)=-\chi\,(y-y')-(1-\chi)\,(y-c)\leq 0 \, ,
\end{align*}
since $y'\leq y$ by Lemma \ref{souk:lemma} and $y\geq c$ on $\Omega_s$, hence $dE_s/dt=-\mathcal{D}_s(t)$ with $\mathcal{D}_s(t)\geq 0$.
\end{proof}

\vskip 0.5cm

The economic meaning is simple: each meeting either reduces the bargaining room of the agents (separation) or removes it from the market (trade). $E_b$ and $E_s$ play the role of the kinetic energy in the Boltzmann equation for granular media, which is dissipated by inelastic collisions \cite{Villani2002}. Their dissipation characterizes the final state (Theorem \ref{souk:longtime}). A related decreasing quantity, the surplus of unrealized mutually beneficial trades, plays a central role in the classical theory of price formation of \cite{InouaSmith2022}.

\subsection{Convergence and characterization of the final state}

\hskip 0.5cm We denote by $\|\cdot\|_{BL}$ the bounded Lipschitz distance between finite measures,
\begin{align*}
\|\mu-\nu\|_{BL}=\sup\left\lbrace \int\phi\,d(\mu-\nu) \, , \, \|\phi\|_\infty+\mathrm{Lip}(\phi)\leq 1\right\rbrace ,
\end{align*}
which metrizes the narrow convergence of nonnegative measures with bounded mass on a compact set.

\vskip 0.5cm

\begin{theorem}[Convergence to an exhausted market]\label{souk:longtime}{\ }\\
Under Assumption \ref{souk:assumption}, let $(f,g)$ be the solution given by Theorem \ref{souk:wellposed}. Then:
\begin{enumerate}
\item There exist $f_\infty\in\mathcal{M}_+(\Omega_b^R)$, $g_\infty\in\mathcal{M}_+(\Omega_s^R)$ and $h_\infty\in\mathcal{M}_+([0,R])$ such that, as $t\to+\infty$, $f(t)\to f_\infty$ and $g(t)\to g_\infty$ narrowly, and $h(t)\to h_\infty$ in total variation. More precisely, for all $t\geq 0$,
\begin{align}
\label{eqn:14}
\|f(t)-f_\infty\|_{BL}\leq\int_t^{+\infty}\left(\mathcal{D}_b+\Gamma\right)ds \quad , \quad \|g(t)-g_\infty\|_{BL}\leq\int_t^{+\infty}\left(\mathcal{D}_s+\Gamma\right)ds \, ,
\end{align}
where the right-hand sides are finite and go to $0$.
\item The masses converge: $N_b(t)\to N_b^\infty$, $N_s(t)\to N_s^\infty$, with $N_b^\infty-N_s^\infty=N_b(0)-N_s(0)$.
\item If $N_b^\infty>0$ and $N_s^\infty>0$, then for all $(x,v)\in\operatorname{supp}f_\infty$ and $(y,c)\in\operatorname{supp}g_\infty$:
\begin{align}
\label{eqn:15}
x\leq y \, , \qquad \text{and} \qquad x<y \ \Longrightarrow \ x=v \ \text{ and } \ y=c \, .
\end{align}
\end{enumerate}
\end{theorem}

\vskip 0.5cm

\begin{proof}
\emph{(i)} Let $\phi$ be a test function on $\Omega_b^R$ with $\|\phi\|_\infty+\mathrm{Lip}(\phi)\leq 1$. Its McShane extension $\tilde\phi(z)=\inf_{w\in\Omega_b^R}\big(\phi(w)+\mathrm{Lip}(\phi)\,|z-w|\big)$, truncated at $\pm\|\phi\|_\infty$, is a bounded continuous function on $\Omega_b$ with the same bounds, which coincides with $\phi$ on $\Omega_b^R$; since $f(t)$ is concentrated on $\Omega_b^R$, we may use it in (\ref{eqn:4}). By (\ref{eqn:4}),
\begin{align*}
\left|\frac{d}{dt}\int \phi\,df(t)\right| = \left|\beta\int_\Theta\Big(\chi\,\big(\phi(x',v)-\phi(x,v)\big)-(1-\chi)\,\phi(x,v)\Big)\,d\pi_t\right| \leq \beta\int_\Theta\chi\,(x'-x)\,d\pi_t+\Gamma(t)\leq\mathcal{D}_b(t)+\Gamma(t) \, ,
\end{align*}
since $|\phi(x',v)-\phi(x,v)|\leq x'-x$ and $|\phi|\leq 1$. By Lemma \ref{souk:dissipation} and Proposition \ref{souk:mass},
\begin{align*}
\int_0^{+\infty}\mathcal{D}_b\,dt\leq E_b(0)\leq R\,N_b(0) \quad , \quad \int_0^{+\infty}\Gamma\,dt\leq N_b(0) \, .
\end{align*}
Hence $\|f(t)-f(s)\|_{BL}\leq\int_s^t(\mathcal{D}_b+\Gamma)\,d\sigma$ for $s\leq t$, and $(f(t))$ is a Cauchy family for the distance $\|\cdot\|_{BL}$. The set of nonnegative measures on the compact set $\Omega_b^R$ with mass at most $N_b(0)$ is compact for the narrow topology (Prokhorov's theorem), which is metrized by $\|\cdot\|_{BL}$; therefore $f(t)$ converges to some $f_\infty$, and letting $t'\to+\infty$ in $\|f(t)-f(t')\|_{BL}\leq\int_t^{t'}(\mathcal{D}_b+\Gamma)\,d\sigma$ gives (\ref{eqn:14}). The same argument applies to $g$. Finally, by Lemma \ref{souk:lemma}, every transaction price $p$ satisfies $c\leq p\leq v\leq R$, so that $h(t)$ is concentrated on $[0,R]$; since $h(t)$ is a nondecreasing family of measures, $\|h(t)-h(s)\|=h(t)([0,R])-h(s)([0,R])$ for $s\leq t$, and its mass $N_b(0)-N_b(t)$ is bounded by $N_b(0)$, so that $h(t)$ converges in total variation to some $h_\infty\in\mathcal{M}_+([0,R])$.\\

\emph{(ii)} follows from Proposition \ref{souk:mass}, since $N_b$ and $N_s$ are nonincreasing and nonnegative.\\

\emph{(iii)} \emph{Step 1: the dissipation vanishes.} By Lemma \ref{souk:lipschitz}, $\|\partial_t f\|$ and $\|\partial_t g\|$ are bounded by $4\beta\left(N_b(0)+N_s(0)\right)^2$, so that $f$ and $g$ are Lipschitz continuous in time for the total variation norm. The integrand $k=\chi\,(x'-x)+(1-\chi)(v-x)+(1-\chi)$ of $\mathcal{D}_b+\Gamma$ is measurable and bounded by $R+1$ on $\Omega_b^R\times\Omega_s^R\times[0,1]^2$, and
\begin{align*}
\|\pi_t-\pi_s\|\leq\|f(t)-f(s)\|\,\|g(t)\|+\|f(s)\|\,\|g(t)-g(s)\| \, ,
\end{align*}
so that $|(\mathcal{D}_b+\Gamma)(t)-(\mathcal{D}_b+\Gamma)(s)|\leq\beta\,(R+1)\,\|\pi_t-\pi_s\|$, and $\mathcal{D}_b+\Gamma$ is Lipschitz continuous in time. A nonnegative, integrable and uniformly continuous function goes to $0$ at infinity: $\mathcal{D}_b(t)+\Gamma(t)\to 0$, and similarly $\mathcal{D}_s(t)+\Gamma(t)\to 0$.\\

\emph{Step 2: a continuous lower bound.} Let
\begin{align*}
H_b(x,v,y,c,a,b)=\min\left(1,(x-y)^+\right)+\min\left(1\, , \, v-x\, , \, a\,(y-x)^+\right) ,
\end{align*}
which is continuous, bounded and nonnegative on $\Theta$. We claim that
\begin{align*}
H_b\leq\chi\,(x'-x)+(1-\chi)(v-x)+(1-\chi) \, .
\end{align*}
If $y\leq x$, then $\chi=0$ and the right-hand side is at least $1\geq H_b$. If $y>x$, then $H_b=\min\left(1,v-x,a(y-x)\right)$; if $\chi=1$, the right-hand side is $x'-x=\min\left(v-x,a(y-x)\right)\geq H_b$; if $\chi=0$, it is at least $v-x\geq H_b$. Consequently,
\begin{align*}
\beta\int_\Theta H_b\,d\pi_t\leq\mathcal{D}_b(t)+\Gamma(t)\xrightarrow[t\to+\infty]{}0 \, .
\end{align*}

\emph{Step 3: passage to the limit.} By (i), $f(t)\to f_\infty$ and $g(t)\to g_\infty$ narrowly, so that $\pi_t=f(t)\otimes g(t)\otimes\Lambda$ converges narrowly to $\pi_\infty=f_\infty\otimes g_\infty\otimes\Lambda$ on the compact set $\Omega_b^R\times\Omega_s^R\times[0,1]^2$ (see \cite{Billingsley1999}). Since $H_b$ is continuous and bounded, $\int H_b\,d\pi_t\to\int H_b\,d\pi_\infty$; since, by Step 2, $0\leq\int H_b\,d\pi_t\leq\beta^{-1}\big(\mathcal{D}_b(t)+\Gamma(t)\big)\to 0$, we obtain $\int H_b\,d\pi_\infty=0$. As $H_b$ is continuous and nonnegative, and as the support of a product of measures is the product of their supports, $H_b$ vanishes on $\operatorname{supp}f_\infty\times\operatorname{supp}g_\infty\times\operatorname{supp}\Lambda$. Let $(x,v)\in\operatorname{supp}f_\infty$ and $(y,c)\in\operatorname{supp}g_\infty$. The first term of $H_b$ gives $x\leq y$. If $x<y$, choosing $(a,b)\in\operatorname{supp}\Lambda$ with $a>0$ (Assumption (A2)), the second term gives $v=x$. The same argument with
\begin{align*}
H_s=\min\left(1,(x-y)^+\right)+\min\left(1\, , \, y-c\, , \, b\,(y-x)^+\right)
\end{align*}
and a point of $\operatorname{supp}\Lambda$ with $b>0$ gives $y=c$.
\end{proof}

\vskip 0.5cm

In words, the souk ends in an \emph{exhausted market}: all the remaining buyers have reached their reservation price, all the remaining sellers have reached theirs, and the remaining offers are below the remaining asking prices. The following corollary translates this into the language of supply and demand.

\vskip 0.5cm

\begin{corollary}[Marshallian bracket]\label{souk:bracket}{\ }\\
Under the assumptions of Theorem \ref{souk:longtime} (iii), let
\begin{align}
\label{eqn:16}
p_-=\max\left\lbrace x \, , \, (x,v)\in\operatorname{supp}f_\infty\right\rbrace \quad , \quad p_+=\min\left\lbrace y \, , \, (y,c)\in\operatorname{supp}g_\infty\right\rbrace ,
\end{align}
and assume that $p_-<p_+$. Then:
\begin{enumerate}
\item every remaining buyer satisfies $v=x\leq p_-$ and every remaining seller satisfies $c=y\geq p_+$, so that no mutually profitable trade remains;
\item every buyer with $v>p_-$ and every seller with $c<p_+$ has traded; in particular, the traded volume $\mathrm{Vol}_\infty=N_b(0)-N_b^\infty$ satisfies, for all $p\in(p_-,p_+)$,
\begin{align}
\label{eqn:17}
\mathrm{Vol}_\infty\geq\max\big(D(p),S(p)\big) \, .
\end{align}
\end{enumerate}
\end{corollary}

\vskip 0.5cm

\begin{proof}
Let $(x,v)\in\operatorname{supp}f_\infty$. Then $x\leq p_-<p_+\leq y$ for every $(y,c)\in\operatorname{supp}g_\infty$, and Theorem \ref{souk:longtime} (iii) gives $x=v$. The same argument gives $y=c$ for the sellers. Let $f^v(t)$ be the marginal of $f(t)$ in $v$. Taking $\phi(x,v)=\theta(v)$, with $\theta\geq 0$, in (\ref{eqn:4}), the integrand is $-(1-\chi)\,\theta(v)\leq 0$: since $v$ is not modified by the meetings, $f^v(t)$ is a nonincreasing family of measures, and $f^v(0)-f^v(t)$ is the distribution of the reservation prices of the buyers who have traded before $t$. Hence $f^v(t)$ converges in total variation, and its limit is the marginal of $f_\infty$, which is concentrated on $\{v\leq p_-\}$. Therefore the buyers with $v>p_-$ have all traded. The same holds for the sellers with $c<p_+$. For $p\in(p_-,p_+)$, $D(p)$ counts buyers with $v\geq p>p_-$, who have all traded, hence $D(p)\leq \mathrm{Vol}_\infty$, and similarly $S(p)\leq \mathrm{Vol}_\infty$.
\end{proof}

\vskip 0.5cm

\begin{remark}{\ }\\
Theorem \ref{souk:longtime} does not exclude the case $p_-=p_+$, where remaining buyers and sellers touch at a common price. The conclusions of Corollary \ref{souk:bracket} then hold for all the agents which are not located at this price. In all our simulations, $p_-<p_+$.
\end{remark}

\vskip 0.5cm

\begin{remark}{\ }\\
Theorem \ref{souk:longtime} and Corollary \ref{souk:bracket} make precise the position of the souk described in the introduction: trades take place before any equilibrium is reached, as in non-t\^atonnement processes \cite{HahnNegishi1962}, and the agents who remain at the end of the day are rationed, as in non-Walrasian equilibria, but the bracket $[p_-,p_+]$ is produced by the bargaining itself, and not by an aggregate price law.
\end{remark}

\vskip 0.5cm

\begin{remark}{\ }\\
By Corollary \ref{souk:bracket}, the Walrasian volume $D(e)$ is a lower bound for the traded volume when $e\in(p_-,p_+)$. In general, $\mathrm{Vol}_\infty$ may exceed $D(e)$: decentralized bargaining allows trades between a buyer with $v<e$ and a seller with $c<e$, or between a buyer with $v>e$ and a seller with $c>e$, which the Walrasian allocation excludes. Whether the bracket $[p_-,p_+]$ contains $e$, and how far the transaction prices are from $e$, depends on the bargaining rule; this question is explored numerically in the next section. For markets with sequential bargaining, results relating decentralized prices and Walrasian prices can be found in \cite{Rubinstein1985,Gale1987}; related kinetic models of price formation can be found in \cite{LasryLions2007,Burger2013}.
\end{remark}

\vskip 0.5cm

\subsection{Expectations and market tension}

\hskip 0.5cm Agents anticipate price movements every day, from the prices they observe, the number of sheep still present, or news from outside the souk. Such expectations change the speed of the concessions: a buyer who expects prices to rise concedes faster, while a seller who expects prices to rise concedes more slowly, and conversely. We do not model how expectations are formed: we first assume that the law of the concession rates depends on time, and then that it depends on the numbers of agents still present. Since expectations change the speed of the concessions but not their direction, buyers still only raise their offers, and sellers only lower their asking prices.

\vskip 0.5cm

\begin{assumption}\label{souk:expect}{\ }
\begin{enumerate}
\item[(A3)] The law of the concession rates is a family $(\Lambda_t)_{t\geq 0}$ of probability measures on $[0,1]^2$, and there exist times $0=t_0<t_1<t_2<\cdots$, finitely many or with $t_k\to+\infty$, such that $t\mapsto\Lambda_t$ is continuous for the total variation norm on each interval $[t_k,t_{k+1})$. Moreover, $\Lambda_t$ converges narrowly, as $t\to+\infty$, to a probability measure $\Lambda_\infty$ which satisfies Assumption (A2).
\end{enumerate}
\end{assumption}

\vskip 0.5cm

\begin{proposition}[Time-dependent expectations]\label{souk:expectprop}{\ }\\
Under Assumptions (A1) and (A3), system (\ref{eqn:4}), with $\pi_t=f(t)\otimes g(t)\otimes\Lambda_t$, has a unique solution, which is nonnegative, and the conclusions of Theorem \ref{souk:longtime} and Corollary \ref{souk:bracket} hold.
\end{proposition}

\vskip 0.5cm

\begin{proof}
\emph{Well-posedness.} By a solution, we mean $F=(f,g)\in C^0(\R_+;X)$ such that (\ref{eqn:4}) holds on each interval $[t_k,t_{k+1})$, with a right derivative at $t_k$. Let $\mathcal{Q}_t$ be the operator (\ref{eqn:9}) with $\Lambda_t$ in place of $\Lambda$. Since $\|\Lambda_t\|=1$, Lemma \ref{souk:lipschitz} holds for $\mathcal{Q}_t$ with the same constants, and
\begin{align*}
\|\mathcal{Q}_t(F)-\mathcal{Q}_s(F)\|_X\leq 2\beta\,\|F\|_X^2\,\|\Lambda_t-\Lambda_s\| \, ,
\end{align*}
so that $(t,F)\mapsto\mathcal{Q}_t(F)$ is continuous on $[t_k,t_{k+1})\times X$, and Lipschitz continuous on bounded sets of $X$, uniformly in $t$. Starting from $F(t_k)$, the proof of Theorem \ref{souk:wellposed} applies without change on $[t_k,t_{k+1})$: the Cauchy-Lipschitz theorem gives a unique solution, which is nonnegative by Step 3, since $\Lambda_t$ is a probability measure, and whose masses are bounded by $N_b(0)+N_s(0)$ by Step 4. Since $\|F'\|_X$ is bounded, $F(t)$ has a limit in $X$ as $t\to t_{k+1}$, which is the initial datum on the next interval. This gives a unique global solution, which is nonnegative, continuous in time, and $C^1$ on each interval $[t_k,t_{k+1})$.\\

\emph{Long-time behaviour.} Proposition \ref{souk:mass}, Proposition \ref{souk:support} and Lemma \ref{souk:dissipation} only use $(a,b)\in[0,1]^2$ and $\Lambda_t([0,1]^2)=1$, and remain valid, the identities of Proposition \ref{souk:mass} and Lemma \ref{souk:dissipation} holding on each interval $[t_k,t_{k+1})$; since $N_b$, $E_b$ and $E_s$ are continuous, we still have $\int_0^{+\infty}\mathcal{D}_b\,dt\leq E_b(0)$ and $\int_0^{+\infty}\Gamma\,dt\leq N_b(0)$. Part (i) of Theorem \ref{souk:longtime} is unchanged. In part (iii), Step 1 is replaced by the following argument: since $\mathcal{D}_b+\Gamma$ is nonnegative and integrable on $\R_+$, there exists a sequence $t_n\to+\infty$ such that $\mathcal{D}_b(t_n)+\Gamma(t_n)\to 0$. By Step 2, $\beta\int_\Theta H_b\,d\pi_{t_n}\to 0$. Since $f(t_n)\to f_\infty$, $g(t_n)\to g_\infty$ and $\Lambda_{t_n}\to\Lambda_\infty$ narrowly, $\pi_{t_n}$ converges narrowly to $f_\infty\otimes g_\infty\otimes\Lambda_\infty$ (see \cite{Billingsley1999}), and Step 3 applies with $\Lambda_\infty$, which satisfies (A2). The same argument holds for $H_s$, along a sequence on which $\mathcal{D}_s+\Gamma$ goes to $0$. The proof of Corollary \ref{souk:bracket} does not depend on $\Lambda$.
\end{proof}

\vskip 0.5cm

Expectations may also respond to the state of the market. A buyer who sees many sheep still unsold concedes more slowly, and a seller who sees the buyers leave concedes faster. The simplest description of this \emph{market tension} is a law of the concession rates which depends on the numbers of agents still present.

\vskip 0.5cm

\begin{assumption}\label{souk:tension}{\ }
\begin{enumerate}
\item[(A4)] The law of the concession rates is $\Lambda[N_b(t),N_s(t)]$, where $(n_b,n_s)\mapsto\Lambda[n_b,n_s]$ is a map from $\R^2$ to the set of probability measures on $[0,1]^2$ such that, for some $L\geq 0$ and all $(n_b,n_s),(m_b,m_s)\in\R^2$,
\begin{align*}
\big\|\Lambda[n_b,n_s]-\Lambda[m_b,m_s]\big\|\leq L\left(|n_b-m_b|+|n_s-m_s|\right) ,
\end{align*}
and $\Lambda[n_b,n_s]$ satisfies Assumption (A2) for all $(n_b,n_s)$.
\end{enumerate}
\end{assumption}

\vskip 0.5cm

\begin{proposition}[Market tension]\label{souk:tensionprop}{\ }\\
Under Assumptions (A1) and (A4), system (\ref{eqn:4}), with $\pi_t=f(t)\otimes g(t)\otimes\Lambda[N_b(t),N_s(t)]$, has a unique solution, which is nonnegative, and the conclusions of Theorem \ref{souk:longtime} and Corollary \ref{souk:bracket} hold.
\end{proposition}

\vskip 0.5cm

\begin{proof}
\emph{Well-posedness.} For $F=(f,g)\in X$, let $n_b(F)=f(\Omega_b)$, $n_s(F)=g(\Omega_s)$, $\Lambda[F]=\Lambda[n_b(F),n_s(F)]$, and let $\mathcal{Q}_{\Lambda}$ denote the operator (\ref{eqn:9}) with the law $\Lambda$. The system reads $F'=\widetilde{\mathcal{Q}}(F)$, with $\widetilde{\mathcal{Q}}(F)=\mathcal{Q}_{\Lambda[F]}(F)$. For $F,G\in X$,
\begin{align*}
\widetilde{\mathcal{Q}}(F)-\widetilde{\mathcal{Q}}(G)=\Big(\mathcal{Q}_{\Lambda[F]}(F)-\mathcal{Q}_{\Lambda[F]}(G)\Big)+\Big(\mathcal{Q}_{\Lambda[F]}(G)-\mathcal{Q}_{\Lambda[G]}(G)\Big) \, .
\end{align*}
By Lemma \ref{souk:lipschitz}, the first term is bounded by $4\beta\left(\|F\|_X+\|G\|_X\right)\|F-G\|_X$. Since the gain terms of (\ref{eqn:9}) are linear in $\Lambda$, with $\|\mathcal{Q}_b^+(f,g)\|\leq\beta\|f\|\,\|g\|\,\|\Lambda\|$, and the loss terms do not depend on $\Lambda$, the second term is bounded by $2\beta\|G\|_X^2\,\|\Lambda[F]-\Lambda[G]\|$. By (A4) and $|n_b(F)-n_b(G)|+|n_s(F)-n_s(G)|\leq\|F-G\|_X$,
\begin{align*}
\big\|\widetilde{\mathcal{Q}}(F)-\widetilde{\mathcal{Q}}(G)\big\|_X\leq\Big(4\beta\left(\|F\|_X+\|G\|_X\right)+2\beta L\,\|G\|_X^2\Big)\|F-G\|_X \, .
\end{align*}
Hence $\widetilde{\mathcal{Q}}$ is Lipschitz continuous on bounded sets of $X$, and Steps 1 and 2 of the proof of Theorem \ref{souk:wellposed} give a unique maximal solution. In Step 3, the masses $n_b(t)$ and $n_s(t)$ of the solution are fixed, and so is the law $\Lambda_t=\Lambda[n_b(t),n_s(t)]$, which is a probability measure: the map $\Psi$ is defined with $\Lambda_t$ in place of $\Lambda$, and the argument applies without change. Step 4 applies as well, so that the solution is global and nonnegative.\\

\emph{Long-time behaviour.} Along the solution, $\Lambda_t=\Lambda[N_b(t),N_s(t)]$ is Lipschitz continuous in time for the total variation norm, since $N_b$ and $N_s$ are Lipschitz continuous by Proposition \ref{souk:mass}. Since $N_b(t)\to N_b^\infty$ and $N_s(t)\to N_s^\infty$, $\Lambda_t$ converges in total variation, hence narrowly, to $\Lambda_\infty=\Lambda[N_b^\infty,N_s^\infty]$, which satisfies (A2). Assumption (A3) thus holds along the solution, with a single interval of continuity, and Proposition \ref{souk:expectprop} gives the conclusions of Theorem \ref{souk:longtime} and Corollary \ref{souk:bracket}.
\end{proof}

\vskip 0.5cm

\begin{remark}{\ }\\
A simple example is a mixture of two laws: $\Lambda[n_b,n_s]=w\,\Lambda_1+(1-w)\,\Lambda_2$, where $\Lambda_1$ describes buyers who concede slowly and sellers who concede fast, $\Lambda_2$ the opposite behaviour, both laws satisfying (A2), and where the weight
\begin{align*}
w=\frac{n_s^+}{n_b^++n_s^++\varepsilon}\in[0,1] \, , \qquad \varepsilon>0 \, ,
\end{align*}
increases with the proportion of sellers still present. Since $w$ is Lipschitz continuous and $\|\Lambda[n]-\Lambda[m]\|\leq 2\,|w(n)-w(m)|$, Assumption (A4) holds. In the basic model, the numbers of agents already act on the prices, through the meeting frequencies $\beta N_s$ and $\beta N_b$ and through the exits; but each agent concedes at each meeting, whatever the number of agents around him, so that a seller who receives many visits concedes more often. A weight $w$ as above lets the agents perceive the imbalance of the market, the side which is the most numerous conceding faster. Concession rates which also depend on the observed prices, or on the history of each agent, would require additional variables, and are left for future work.
\end{remark}

\vskip 0.5cm

\subsection{Agents who trade several sheep}

\hskip 0.5cm On a real souk, a breeder often brings several sheep, and a buyer may buy several sheep, for instance for his family. In this subsection, each agent is also described by his \emph{stock}, i.e.\ the number of sheep he still wants to buy or to sell: a buyer by $(x,v,m)\in\Omega_b^K:=\Omega_b\times\{1,\dots,K\}$, and a seller by $(y,c,k)\in\Omega_s^K:=\Omega_s\times\{1,\dots,K\}$, where the integer $K\geq 1$ is the largest number of sheep that an agent wants to buy or to sell. The same bound is used for buyers and sellers without loss of generality: if buyers want at most $K_b$ sheep and sellers bring at most $K_s$ sheep, one takes $K=\max(K_b,K_s)$, and the measures give no mass to the larger stocks. For simplicity, an agent has the same reservation price for all his sheep. The distributions $f(t)$ and $g(t)$ are nonnegative measures on $\Omega_b^K$ and $\Omega_s^K$, and we distinguish the numbers of agents and the numbers of sheep still demanded and offered,
\begin{align*}
N_b(t)=f(t)(\Omega_b^K) \, , \quad N_s(t)=g(t)(\Omega_s^K) \, , \quad M_b(t)=\int m\,df(t) \, , \quad M_s(t)=\int k\,dg(t) \, .
\end{align*}
The Marshall curves now count sheep: $D(p)=\int\1_{\{v\geq p\}}\,m\,df_0$ and $S(p)=\int\1_{\{c\leq p\}}\,k\,dg_0$. The bargaining rule (\ref{eqn:2}) and the indicator $\chi$ of a separation are unchanged, and do not depend on the stocks. After a trade, one sheep changes hands: the buyer moves to $(x,v,m-1)$ if $m\geq 2$, and leaves the market if $m=1$; the seller moves to $(y,c,k-1)$ if $k\geq 2$, and leaves the market if $k=1$. Both agents start the bargaining for their next sheep from the prices $x$ and $y$ they posted before the trade. With $\Theta^K=\Omega_b^K\times\Omega_s^K\times[0,1]^2$ and $\pi_t=f(t)\otimes g(t)\otimes\Lambda$, the evolution of $f$ and $g$ is given, for all bounded continuous test functions $\phi$ on $\Omega_b^K$ and $\psi$ on $\Omega_s^K$, by
\begin{align}
\label{eqn:18}
\begin{cases}
\displaystyle\frac{d}{dt}\int\phi\,df(t)=\beta\int_{\Theta^K}\Big(\chi\,\phi(x',v,m)+(1-\chi)\,\1_{\{m\geq 2\}}\,\phi(x,v,m-1)-\phi(x,v,m)\Big)\,d\pi_t \\ \\
\displaystyle\frac{d}{dt}\int\psi\,dg(t)=\beta\int_{\Theta^K}\Big(\chi\,\psi(y',c,k)+(1-\chi)\,\1_{\{k\geq 2\}}\,\psi(y,c,k-1)-\psi(y,c,k)\Big)\,d\pi_t
\end{cases}
\end{align}
For $K=1$, we recover system (\ref{eqn:4}).

\vskip 0.5cm

\begin{assumption}\label{souk:stock}{\ }
\begin{enumerate}
\item[(A5)] The stocks are bounded by $K$, $f_0$ and $g_0$ are nonnegative measures concentrated on $\Omega_b^R\times\{1,\dots,K\}$ and $\Omega_s^R\times\{1,\dots,K\}$, and $\Lambda$ satisfies Assumption (A2).
\end{enumerate}
\end{assumption}

\vskip 0.5cm

\begin{proposition}[Agents who trade several sheep]\label{souk:stockprop}{\ }\\
Under Assumption (A5), system (\ref{eqn:18}) has a unique solution $(f,g)\in C^1\big(\R_+;\mathcal{M}(\Omega_b^K)\times\mathcal{M}(\Omega_s^K)\big)$, which is nonnegative. Moreover:
\begin{enumerate}
\item sheep demanded and sheep offered leave the market by pairs:
\begin{align*}
\frac{dM_b}{dt}=\frac{dM_s}{dt}=-\,\Gamma(t) \quad , \quad M_b(t)-M_s(t)=M_b(0)-M_s(0) \, ,
\end{align*}
while the numbers of agents $N_b$ and $N_s$ are nonincreasing;
\item the bargaining rooms $E_b(t)=\int m\,(v-x)\,df(t)$ and $E_s(t)=\int k\,(y-c)\,dg(t)$ are nonincreasing;
\item the conclusions of Theorem \ref{souk:longtime} hold, and so do those of Corollary \ref{souk:bracket}, with $\mathrm{Vol}_\infty=M_b(0)-M_b^\infty=M_s(0)-M_s^\infty$ and with Marshall curves counting sheep; in particular, every buyer with $v>p_-$ has bought all his sheep, and every seller with $c<p_+$ has sold all his sheep.
\end{enumerate}
\end{proposition}

\vskip 0.5cm

\begin{proof}
\emph{Well-posedness.} The gain term of the buyers is the image of $\beta\,\chi\,\pi_t$ by the map $(x,v,m,y,c,k,a,b)\mapsto(x',v,m)$, plus the image of $\beta\,(1-\chi)\,\1_{\{m\geq 2\}}\,\pi_t$ by the map $(x,v,m,y,c,k,a,b)\mapsto(x,v,m-1)$. These two measures are carried by disjoint events, so that the total variation of their sum is at most $\beta\|f\|\,\|g\|$, and the loss term is $\beta\,g(\Omega_s^K)\,f$, as in (\ref{eqn:9}). The same holds for the sellers. Hence Lemma \ref{souk:lipschitz} holds with the same constants, and the proofs of Theorem \ref{souk:wellposed} and of Proposition \ref{souk:support} apply without change in $\mathcal{M}(\Omega_b^K)\times\mathcal{M}(\Omega_s^K)$.\\

\emph{(i)} With $\phi=m$, which is bounded by (A5), the integrand of the first line of (\ref{eqn:18}) is $\chi m+(1-\chi)(m-1)-m=-(1-\chi)$ if $m\geq 2$, and $\chi-1=-(1-\chi)$ if $m=1$, hence $dM_b/dt=-\Gamma$; the same computation with $\psi=k$ gives $dM_s/dt=-\Gamma$. With $\phi=1$ and $\psi=1$, the integrands are $-(1-\chi)\1_{\{m=1\}}\leq 0$ and $-(1-\chi)\1_{\{k=1\}}\leq 0$.\\

\emph{(ii)} With $\phi=m\,\min(v-x,R)$, which coincides with $m(v-x)$ on $\Omega_b^R\times\{1,\dots,K\}$, the integrand of the first line of (\ref{eqn:18}) is $-m(x'-x)\leq 0$ in case of separation, $(m-1)(v-x)-m(v-x)=-(v-x)\leq 0$ in case of a trade with $m\geq 2$, and $-(v-x)\leq 0$ in case of a trade with $m=1$. Hence $dE_b/dt=-\mathcal{D}_b^K$, with
\begin{align*}
\mathcal{D}_b^K(t)=\beta\int_{\Theta^K}\Big(\chi\,m\,(x'-x)+(1-\chi)\,(v-x)\Big)\,d\pi_t\geq 0 \, ,
\end{align*}
and the same computation with $\psi=k\,\min(y-c,R)$ gives $dE_s/dt=-\mathcal{D}_s^K$, with
\begin{align*}
\mathcal{D}_s^K(t)=\beta\int_{\Theta^K}\Big(\chi\,k\,(y-y')+(1-\chi)\,(y-c)\Big)\,d\pi_t\geq 0 \, .
\end{align*}

\emph{(iii)} We endow $\Omega_b^K$ with the distance $|(x,v)-(\tilde x,\tilde v)|+|m-\tilde m|$, and $\Omega_s^K$ with the analogous distance. If $\|\phi\|_\infty+\mathrm{Lip}(\phi)\leq 1$, the integrand of the first line of (\ref{eqn:18}) is bounded by $\chi\,(x'-x)+2(1-\chi)$, which is bounded by the integrand of $\mathcal{D}_b^K$ plus $2(1-\chi)$, since $m\geq 1$; the same holds for the sellers. Since $\int_0^{+\infty}\mathcal{D}_b^K\,dt\leq E_b(0)\leq KR\,N_b(0)$, $\int_0^{+\infty}\mathcal{D}_s^K\,dt\leq E_s(0)\leq KR\,N_s(0)$ and $\int_0^{+\infty}\Gamma\,dt\leq M_b(0)$, part (i) of Theorem \ref{souk:longtime} holds on the compact sets $\Omega_b^R\times\{1,\dots,K\}$ and $\Omega_s^R\times\{1,\dots,K\}$, and part (ii) follows from item (i). In part (iii), Step 1 holds with $\mathcal{D}_b^K$ and $\mathcal{D}_s^K$, whose integrands are bounded by $KR$, and Step 2 holds with the same functions $H_b$ and $H_s$, since, if $y>x$ and $\chi=1$, $m(x'-x)\geq x'-x\geq H_b$ and $k(y-y')\geq y-y'\geq H_s$, and the other cases are unchanged. Step 3 applies without change. For Corollary \ref{souk:bracket}, taking $\phi=m\,\theta(v)$ and $\psi=k\,\theta(c)$, with $\theta\geq 0$, in (\ref{eqn:18}), the integrands are $0$ in case of separation and $-\theta(v)$, $-\theta(c)$ in case of a trade: the measures $\int m\,\1_{\{v\in\cdot\}}\,df(t)$ and $\int k\,\1_{\{c\in\cdot\}}\,dg(t)$, which count the sheep still demanded and offered according to the reservation prices of their owners, are nonincreasing, and converge in total variation to their values for $f_\infty$ and $g_\infty$, which are concentrated on $\{v\leq p_-\}$ and on $\{c\geq p_+\}$. Hence every buyer with $v>p_-$ has bought all his sheep, and every seller with $c<p_+$ has sold all his sheep, so that $D(p)\leq\mathrm{Vol}_\infty$ for $p>p_-$ and $S(p)\leq\mathrm{Vol}_\infty$ for $p<p_+$.
\end{proof}

\vskip 0.5cm

\begin{remark}{\ }\\
After a trade, the agents start the bargaining for their next sheep from their prices before the trade, so that the concessions made in a successful bargaining are not kept. This is a modelling choice: any measurable rules $(\hat x,\hat y)$ for the prices after a trade with $x\leq\hat x\leq v$ and $c\leq\hat y\leq y$ lead to the same results, with the same proof. For instance, the agents may keep their concessions, $(\hat x,\hat y)=(x',y')$, when the trade follows a concession; the rule $(x',y')$ cannot be used when $y\leq x$, since $x'\leq x$ and $y'\geq y$ in this case. A buyer who has just bought a sheep may also lower his offer for the next one, and a seller who sells quickly may raise his asking price; such rules would require a different Lyapunov functional. Likewise, a buyer may value his second sheep less than his first: this would require a reservation price for each sheep, and is left for future work.
\end{remark}

\vskip 0.5cm

\section{Numerical illustration\label{Section4}}

\hskip 0.5cm We simulate the souk with a particle method, which is the microscopic counterpart of (\ref{eqn:4}), in the spirit of the Direct Simulation Monte Carlo (DSMC) method of Bird \cite{bird1994,Pareschi2001}.\\

For a time step $\Delta t$, the explicit time discretization of (\ref{eqn:4}) reads
\begin{align}
\label{eqn:19}
\begin{cases}
f^{n+1}=\left(1-\beta\,\Delta t\,N_s^n\right)f^n+\Delta t\,\mathcal{Q}_b^+(f^n,g^n) \, , \quad f^0=f_0 \, ,\\
g^{n+1}=\left(1-\beta\,\Delta t\,N_b^n\right)g^n+\Delta t\,\mathcal{Q}_s^+(f^n,g^n) \, , \quad g^0=g_0 \, ,
\end{cases}
\end{align}
where $N_b^n$ and $N_s^n$ are the masses of $f^n$ and $g^n$. It is interpreted as follows: during $\Delta t$, a buyer does not meet any seller with probability $1-\beta\,\Delta t\,N_s^n$, and meets a seller with probability $\beta\,\Delta t\,N_s^n$; the meeting then ends with a trade, and the buyer leaves the market, or with a separation, and the buyer moves to his new offer $x'$. With $n_b$ buyers and $n_s$ sellers represented by particles, and masses normalized by $N$, the number of meetings during $\Delta t$ is $N_c=\beta\,\Delta t\,n_b\,n_s/N$. The resulting algorithm is the following.\\

\begin{center}
\begin{tabular}{ll}
\hline
\multicolumn{2}{l}{\textbf{Algorithm} (DSMC for the souk)}\\
\hline
\emph{Input:} & a time step $\Delta t$, a number of steps $M$, the numbers $N$ of buyers and sellers.\\
\emph{Initialization:} & sample the lists $(x_i,v_i)$ and $(y_j,c_j)$ from $f_0$ and $g_0$.\\
\emph{For} $n=1$ to $M$: & compute $N_c=\beta\,\Delta t\,n_b\,n_s/N$ and select $N_c$ random pairs $(i,j)$\\
& of a buyer and a seller still present; for each pair, draw $(a,b)$ from $\Lambda$\\
& and apply rule (\ref{eqn:2}): remove both agents in case of trade, and record\\
& the transaction price; otherwise, set $x_i=x'$ and $y_j=y'$.\\
\emph{Output:} & the remaining agents and the list of transaction prices.\\
\hline
\end{tabular}
\end{center}

In practice, we use this algorithm with $\beta=1$, $\Delta t=0.01$ for $t<20$ and $\Delta t=0.1$ afterwards; the non-integer part of $N_c$ is treated by a random rounding. All the simulations below are performed with one sheep per agent, i.e.\ with $K=1$ in the setting of Section \ref{Section3}; the algorithm extends to agents who trade several sheep by decreasing their stocks after each trade, instead of removing them.\\

We consider $N=2000$ buyers and $N=2000$ sellers, with reservation prices drawn from normal distributions, in Moroccan dirhams (MAD):
\begin{align*}
v\sim\mathcal{N}(3500,600^2) \quad , \quad c\sim\mathcal{N}(3000,600^2) \, ,
\end{align*}
truncated below at $500$. Buyers open below their reservation price and sellers above theirs: $x_0=v\,(1-u)$ and $y_0=c\,(1+w)$, with $u\sim\mathcal{U}([0.15,0.35])$ and $w\sim\mathcal{U}([0.15,0.45])$. The concession rates are drawn from the uniform distribution $\Lambda$ on $[0,\bar a]\times[0,\bar b]$. The simulation stops when the market is exhausted, in the sense of Theorem \ref{souk:longtime}. The Walrasian price and volume are computed from the empirical curves (\ref{eqn:1}).\\

\begin{figure}[h!]
\centering
\begin{subfigure}[t]{0.49\textwidth}
\centering
\includegraphics[width=\textwidth]{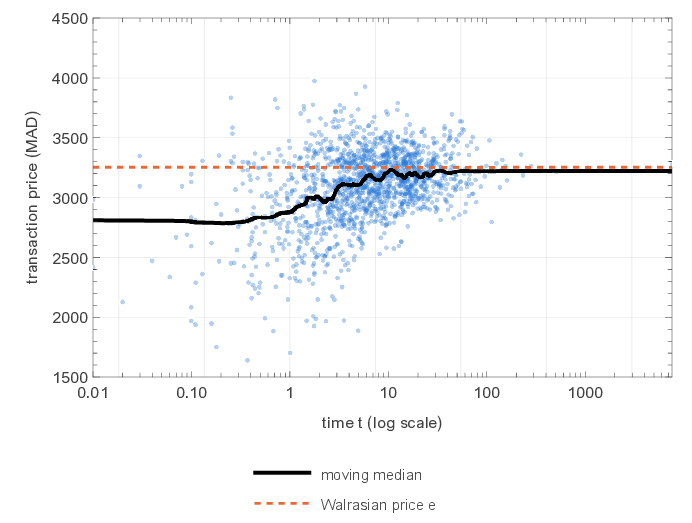}
\caption{Transaction prices and their smoothed moving median (101 trades).}
\end{subfigure}
\hfill
\begin{subfigure}[t]{0.49\textwidth}
\centering
\includegraphics[width=\textwidth]{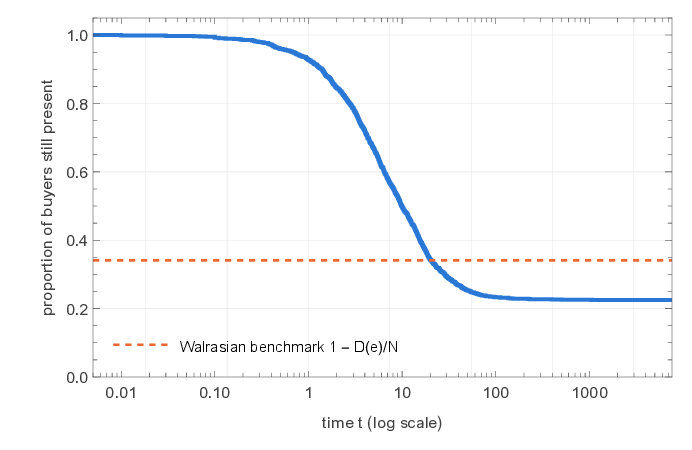}
\caption{Proportion of buyers still present.}
\end{subfigure}
\caption{Reference run ($\bar a=\bar b=0.3$): dynamics of the souk.}
\label{fig:dynamics}
\end{figure}

\begin{figure}[h!]
\centering
\includegraphics[width=0.6\textwidth]{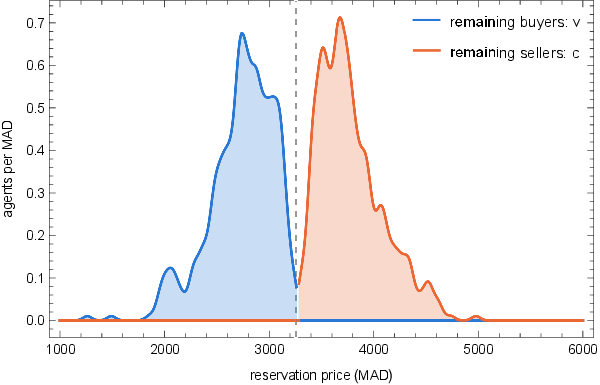}
\caption{Reference run: smoothed distributions of the reservation prices of the agents still present at the end of the day, which lie below $p_-$ for the buyers and above $p_+$ for the sellers. The dashed line is the Walrasian price $e$.}
\label{fig:final}
\end{figure}

Figures \ref{fig:dynamics} and \ref{fig:final} show a reference run with $\bar a=\bar b=0.3$, for which $e=3253$ MAD and $D(e)=1318$. Since each meeting involves one buyer, time can be measured in meetings per buyer: in this run, half of the trades are realized after $3.9$ meetings per buyer on average, $95\%$ after $8.8$ and $99\%$ after $12.8$. The exhausted state of Theorem \ref{souk:longtime} is thus nearly reached within a realistic number of visits during a souk day.

We observe the following.
\begin{enumerate}
\item The first trades take place at low prices. The median transaction price then increases and stabilizes close to $e$ (Figure \ref{fig:dynamics}a). This is a selection effect: the sellers with low costs announce low prices and sell first. Splitting the trades into three groups by time, the mean cost of the sellers is $2457$, $2843$ and $3072$ MAD, and the mean price is $315$, $112$ and $54$ MAD below $e$.
\item The number of buyers decreases to a positive limit (Figure \ref{fig:dynamics}b). The final number of trades, $1550$, exceeds the Walrasian volume $1318$.
\item At the end of the day, the remaining buyers and sellers are separated (Figure \ref{fig:final}): $p_-=3258$ MAD and $p_+=3282$ MAD, as predicted by Corollary \ref{souk:bracket}. In this run, the bracket $[p_-,p_+]$ lies just above the Walrasian price $e=3253$ MAD, within $29$ MAD of it.
\end{enumerate}

\begin{figure}[h!]
\centering
\includegraphics[width=0.9\textwidth]{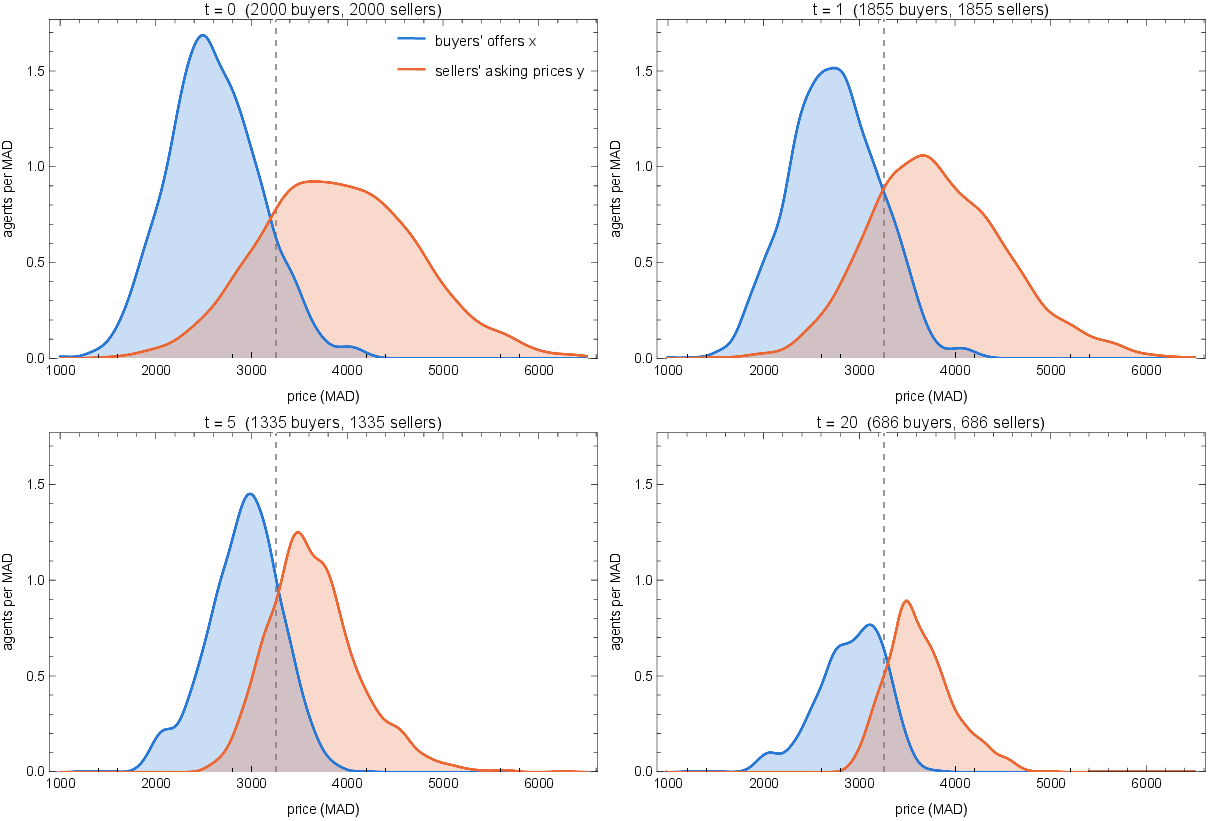}
\caption{Smoothed distributions of the offers $x$ of the buyers and of the asking prices $y$ of the sellers still present at $t=0,1,5,20$ ($\bar a=\bar b=0.3$). The dashed line is the Walrasian price $e$.}
\label{fig:distributions}
\end{figure}

Figure \ref{fig:distributions} shows the offers of the buyers and the asking prices of the sellers still present at several times, for the same run. The offers move up and the asking prices move down, the two distributions lose mass by pairs, and they end up separated on both sides of $e$.

Table \ref{tab:sensitivity} reports averages over $5$ runs for several concession rates.

\begin{table}[h!]
\centering
\begin{tabular}{ccccc}
\hline
$\bar a$ & $\bar b$ & $\mathrm{Vol}_\infty/D(e)$ & mean price $-\,e$ (MAD) & $[p_- - e\, , \, p_+ - e]$ (MAD) \\
\hline
$0.01$ & $0.01$ & $1.04$ & $-117\pm8$ & $[-13\, , \,-9]$ \\
$0.05$ & $0.05$ & $1.07$ & $-125\pm8$ & $[-19\, , \,-10]$ \\
$0.30$ & $0.30$ & $1.16$ & $-158\pm6$ & $[-16\, , \,+6]$ \\
$0.80$ & $0.80$ & $1.23$ & $-175\pm8$ & $[-21\, , \,+14]$ \\
$0.30$ & $0.05$ & $1.10$ & $+23\pm8$ & $[-22\, , \,-2]$ \\
$0.05$ & $0.30$ & $1.09$ & $-312\pm7$ & $[+17\, , \,+29]$ \\
\hline
\end{tabular}
\caption{Sensitivity to the concession rates: averages over 5 runs, with the standard deviation across runs for the mean price. The standard deviation across runs is at most $0.01$ for $\mathrm{Vol}_\infty/D(e)$ and at most $29$ MAD for the bounds of the bracket. The standard deviation of the reservation prices is 600 MAD.}
\label{tab:sensitivity}
\end{table}

\begin{enumerate}
\item In all cases, with equal numbers of buyers and sellers, the bracket $[p_-,p_+]$ lies within $30$ MAD of $e$, i.e.\ less than one tenth of the standard deviation of the reservation prices: the final state of the souk is close to the Marshallian equilibrium.
\item The number of trades exceeds the Walrasian volume by $4\%$ to $23\%$, and the excess increases with the concession rates: fast bargaining produces more trades between agents that the Walrasian allocation would exclude. This excess of trades, together with transaction prices below $e$, was already observed by Chamberlin in decentralized classroom markets \cite{Chamberlin1948}.
\item The mean transaction price depends on the bargaining power of the two sides: when buyers concede faster than sellers ($\bar a>\bar b$), prices are higher, and conversely. With symmetric rates, the mean price remains below $e$, because of the selection effect described above. This bias is not due to the asymmetric opening margins: with symmetric margins ($u,w\sim\mathcal{U}([0.15,0.35])$), the mean price is still $205$ MAD below $e$ for $\bar a=\bar b=0.3$, and $187$ MAD below $e$ for $\bar a=\bar b=0.05$.
\end{enumerate}

When the concession rates decrease from $0.8$ to $0.01$, the ratio $\mathrm{Vol}_\infty/D(e)$ decreases from $1.23$ to $1.04$, and for $\bar a=\bar b=0.01$ the bracket lies within $13$ MAD of $e$. These observations suggest the following conjecture, which we leave for future work.

\vskip 0.5cm

\begin{conjecture}{\ }\\
Assume that $N_b(0)=N_s(0)$. When the concession rates go to $0$, the bracket $[p_-,p_+]$ shrinks to the Walrasian price $e$, and $\mathrm{Vol}_\infty$ converges to the Walrasian volume $D(e)$.
\end{conjecture}

\vskip 0.5cm

\begin{figure}[h!]
\centering
\includegraphics[width=0.6\textwidth]{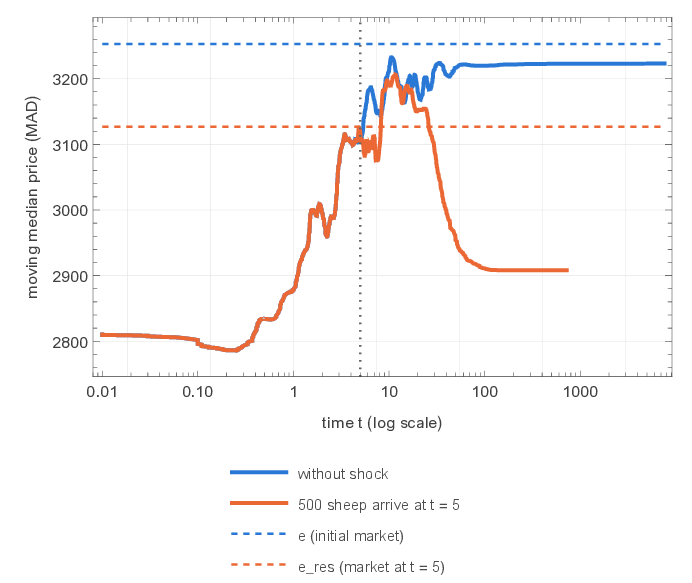}
\caption{Smoothed moving median of the transaction prices (101 trades), without shock and with $500$ additional sheep arriving at $t=5$ (dotted line). The dashed lines are the Walrasian prices of the initial market and of the market present at $t=5$.}
\label{fig:shock}
\end{figure}

Finally, we study the effect of a shock. In the reference run, a truck of $500$ sheep arrives at $t=5$, when $1335$ buyers and $1335$ sellers are still present (Figure \ref{fig:distributions}): we add $500$ sellers, drawn from the same distribution as the initial ones. The Walrasian price of the market present at $t=5$, including the new sellers, is $e_{res}=3127$ MAD, against $e=3253$ MAD for the initial market. Figure \ref{fig:shock} compares the transaction prices with and without the shock. Prices drop after the arrival of the sheep, and the final bracket is $[2987,2992]$ MAD, about $137$ MAD below $e_{res}$; there are $1704$ trades instead of $1550$, $796$ unsold sheep and $296$ buyers without a sheep. Over $3$ runs, the final bracket lies $84$ to $145$ MAD below $e_{res}$. The final state is driven by the new supply, but it lies below the Walrasian price of the residual market: the timing of the arrivals matters.

\vskip 0.5cm

\begin{figure}[h!]
\centering
\includegraphics[width=0.7\textwidth]{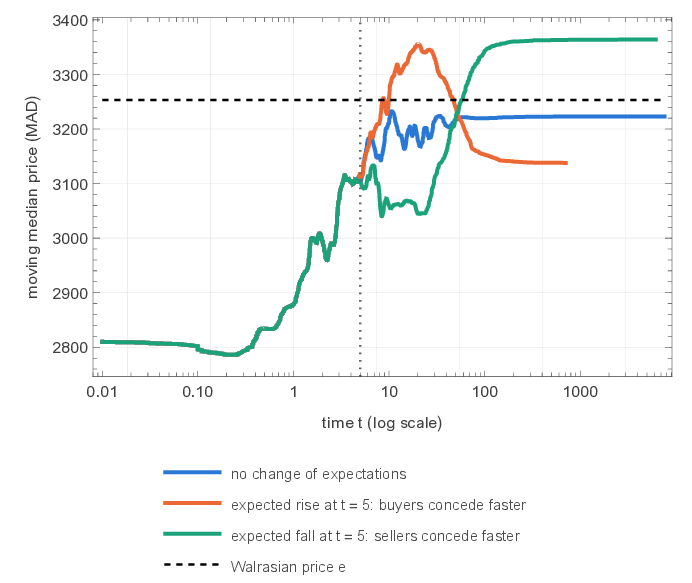}
\caption{Smoothed moving median of the transaction prices (101 trades) when expectations change at $t=5$ (dotted line): no change ($\bar a=\bar b=0.3$), expected rise ($\bar a=0.6$, $\bar b=0.1$) and expected fall ($\bar a=0.1$, $\bar b=0.6$). The dashed line is the Walrasian price $e$.}
\label{fig:expectations}
\end{figure}

Figure \ref{fig:expectations} illustrates the effect of a change of expectations at $t=5$, in the reference run, in the setting of Section \ref{Section3}. Before $t=5$, both sides concede at the same pace, with $\bar a=\bar b=0.3$. After $t=5$, a rumour of a price rise makes buyers concede faster and sellers more slowly ($\bar a=0.6$, $\bar b=0.1$), while a rumour of a price fall has the opposite effect ($\bar a=0.1$, $\bar b=0.6$). The rumour first fulfils itself: averaged over $3$ runs, the trades realized after $t=5$ take place about $78$ MAD above the reference run in the case of an expected rise, and about $67$ MAD below it in the case of an expected fall. The effect is then reversed at the end of the day: the final bracket lies $38$ to $79$ MAD below $e$ after an expected rise, and $41$ to $120$ MAD above $e$ after an expected fall, while the traded volume is almost unchanged, with $\mathrm{Vol}_\infty/D(e)$ between $1.13$ and $1.18$ in all cases. Expectations thus move prices during the day in the direction of the rumour, but they do not change the volume of trade.

\vskip 0.5cm

\section{Conclusion}

\hskip 0.5cm This paper proposes an explicit price adjustment mechanism for bargaining markets, built on the bargaining observed on the sheep souks before Eid al-Adha. Buyers and sellers meet at random, observe the gap between their prices, concede part of it, and trade when the gap is closed. The resulting kinetic equation is a Boltzmann-type equation with losses: buyers and sellers leave the market by pairs.\\

We proved that this equation is well-posed in the space of measures, and that the market converges to an exhausted state, in which no mutually profitable trade remains: the remaining buyers have reached their reservation price, the remaining sellers too, and a price bracket separates them. In simulations, this state is nearly reached after about ten meetings per buyer. The supply and demand curves are not postulated but derived from the distributions of reservation prices, and the bracket provides a decentralized counterpart of Marshall's scissors. Numerical simulations show that this bracket lies close to the Walrasian price, while decentralized bargaining produces more trades than the Walrasian allocation and transaction prices that depend on the bargaining power of each side. After a supply shock during the day, the souk ends below the Walrasian price of the residual market, which shows that the timing of arrivals matters. The theoretical results extend to expectations that vary during the day, or respond to the number of agents still present, and to agents who buy or sell several sheep, the sheep demanded and offered then leaving the market by pairs; simulations show that a rumour of a price rise or fall moves the prices during the day in its direction, but does not change the volume of trade.\\

Contrary to the Walrasian t\^atonnement, trades take place out of equilibrium, and the limit price is not fixed by the opening prices. This approach addresses, for a single market, the criticism of Mark Blaug \cite{Blaug1992}, for whom economists remained ``almost as far away as Walras was'' from identifying the real-world forces that bring markets to equilibrium. Although the Eid souk is our guiding example, the model applies to bargaining markets in general, such as bazaars, livestock markets, flea markets or second-hand markets between individuals. It does not take into account the repeated relationships between buyers and sellers, or clientelization, which Geertz identifies as a central feature of bazaars \cite{Geertz1978}, nor the heterogeneity of the goods.\\

Several extensions are natural: linking the weekly souks until Eid, with unsold sheep and unsatisfied buyers coming back the following week; considering markets where supply and demand are renewed continuously, such as vegetable markets, by adding entry terms to (\ref{eqn:4}); taking into account the heterogeneity of the goods, such as the weight or the breed of the sheep, and the repeated relationships between buyers and sellers; letting expectations depend on the observed prices, on the history of each agent and on outside information; introducing rational expectations, in the framework of mean field games \cite{LasryLions2007}, where agents may delay their purchases in anticipation of lower prices; simulating markets with unequal numbers of buyers and sellers, and agents who trade several sheep; proving the conjecture of Section \ref{Section4}; calibrating the model on observed prices; and, following Smale's eighth problem \cite{Smale1998}, extending the mechanism to several interdependent markets, where the complementarity between goods is expected to be the main difficulty.\\

\section*{Declarations}

\subsection*{Conflict of interest}

The authors declare no conflict of interest.

\subsection*{Authors' contributions}

Both authors contributed to the conception of the model, the mathematical analysis, the numerical simulations and the writing of the manuscript. Both authors read and approved the final manuscript.

\subsection*{Funding}

No funding was received for conducting this study.

\subsection*{Availability of data and materials}

No external data were used. The numerical examples are generated from the parameters given in Section \ref{Section4}; the computations were carried out with Wolfram Mathematica, and the code is available from the authors upon request.

\subsection*{Declaration of generative AI and AI-assisted technologies in the writing process}

During the preparation of this work, the authors used Claude (Anthropic) to review the manuscript, suggest additional references, assist in the writing of the proofs of the extensions of the model proposed by the authors, and check the numerical simulations, which were carried out by the authors with Wolfram Mathematica. After using this tool, the authors reviewed, verified and edited all the content, and take full responsibility for the content of the publication.

\bibliographystyle{apalike}
\bibliography{BibliographieEco}

\end{document}